\documentclass[journal,twoside,web]{IEEEtran}
\usepackage{cite}
\usepackage{amsmath,amssymb,amsfonts}
\usepackage{bm}
\usepackage{algorithm}
\usepackage{algorithmic}
\usepackage{graphicx}
\usepackage{epstopdf}
\usepackage{booktabs}
\usepackage{multirow}
\usepackage{textcomp}
\usepackage{subfigure}
\usepackage{hyperref}
\hypersetup{hidelinks}
\usepackage{cases}
\newtheorem{proof}{Proof}
\newtheorem{property}{Property}
\usepackage{color,xcolor}

\begin{document}
\title{Multi-UAV Mobile Edge Computing and Path Planning Platform based on Reinforcement Learning}

\author{Huan Chang$^{\dag}$\thanks{Huan Chang is with Beihang University, Beijing, China; The civil engineering department, Empire College London}, Yicheng Chen$^{\dag}$\thanks{Yicheng Chen is with Beihang University, Beijing, China}, Baochang Zhang* \thanks{Baochang Zhang is with Beihang University, Beijing, China} \IEEEmembership{Senior Member, IEEE}, David Doermann\thanks{David Doermann is with University at Buffalo, USA} \IEEEmembership{Fellow, IEEE}\thanks{$^{\dag}$ Huan Chang and Yicheng Chen contributed equally to this paper.}\thanks{* Corresponding Author}\thanks{\copyright~2021 IEEE. Personal use of this material is permitted. Permission from IEEE must be obtained for all other uses, in any current or future media, including reprinting/republishing this material for advertising or promotional purposes, creating new collective works, for resale or redistribution to servers or lists, or reuse of any copyrighted component of this work in other works.}}

\maketitle

\begin{abstract}
Unmanned Aerial vehicles (UAVs) are widely used as network processors in mobile networks, but more recently, UAVs have been used in Mobile Edge Computing as mobile servers. However, there are significant challenges to use UAVs in complex environments with obstacles and cooperation between UAVs. We introduce a new multi-UAV Mobile Edge Computing platform, which aims to provide better Quality-of-Service and path planning based on reinforcement learning to address these issues. The contributions of our work include: 1) optimizing the quality of service for mobile edge computing and path planning in the same reinforcement learning framework; 2) using a sigmoid-like function to depict the terminal users' demand to ensure a higher quality of service; 3) applying synthetic considerations of the terminal users' demand, risk and geometric distance in reinforcement learning reward matrix to ensure the quality of service, risk avoidance, and the cost-savings. Simulations have shown the effectiveness and feasibility of our platform, which can help advance related researches. The source code can be found at \url{https://github.com/bczhangbczhang}.
\end{abstract}


\section{Introduction}
\label{sec:introduction}
\IEEEPARstart{M}{obile} data processing technology is experiencing a growing demand in the communication market.
New technologies like 5G are emerging to accelerate its development. However, the demands of terminal users in uncertain environments and extreme situations have never been perfectly satisfied as the calculations and services are often hard to reach from base stations. As a result, mobile edge computing comes out as one of the fastest-growing topics in telecommunications in the past few decades \cite{abbas2017mobile}.

Mobile edge computing is a concept that integrates the capabilities of the network, computing, storage, and intelligence services on the edge of the network physically close to the data source. In a typical mobile edge computing scenario, terminal users are served by edge servers with high computing power \cite{mao2017survey,mach2017mobile}. The effectiveness of mobile edge computing is measured by the Quality of Service ($QoS$) of each terminal user. The higher the $QoS$, the more efficient a terminal user’s demand is satisfied or served. 

Unmanned Aerial Vehicles (UAVs) have become ideal servers for mobile edge computing that assures $QoS$,  improving stability, reliability, and calculating efficiency through research and development investment  \cite{du2019joint,zeng2016wireless,gupta2015survey} . They are also flexible and cost-effective due to their small size \cite{yang2020multi,luo2020optimization}. Therefore, a UAV can move flexibly from terminal user to terminal user and conduct highly efficient calculating services to improve $QoS$.

UAV-mounted mobile edge computing remains challenging due to the complexity of the working environment, the uncertainty of the distribution of terminal users, and the limitations of the UAV’s energy \cite{abbas2017mobile}. Therefore, path planning plays an indispensable role when using UAVs for mobile edge computing to handle these issues. 
For example, \cite{liu2019trajectory} proposed a multi-agent algorithm to determine the optimal path of UAVs based on Q-learning. An echo state network (ESN) based prediction algorithm is also proposed for predicting the future movement of terminal users. Liu et al. \cite{liu2020path} built a model to evaluate $QoS$ and proposed an algorithm to maximize the reward during the planning process. 

However, most of the existing research focuses on finding the optimal path under a specified mission or treating the planning process as a simple greedy strategy without a continuously improving process. This prevents the UAVs’ adaption to the changing environment, and the planning can easily fall into a local optimum. Furthermore, previous works on UAV-mounted mobile edge computing seldom consider risk avoidance or collisions between UAVs as this is impractical in real environments.

To find an adaptive global optimum solution for different tasks, \cite{faraci2019reinforcement,huang2019deep} proves that Reinforcement Learning (RL) is effective. \cite{wang2019multi} dramatically improved $QoS$ by introducing RL and regards the path planning problem as an optimization problem with constraints from the environment. Compared with traditional path planning methods such as the A* algorithm and RRT, RL is more flexible for the following reasons: 1) In the mobile edge computing scenario, the terminal user's residual demand dynamically changes, which requires real-time policy updating. However, traditional methods cannot work efficiently in such time-varying scenarios; 2) There are both obstacles and terminal users on the map, which requires not only obstacle avoidance but also task allocation, and thus it is difficult for those algorithms that only consider geometrical constraints to handle this issue; 3) In RL, elements in the environment can be depicted uniformly by the cost function, thus making it possible to flexibly adjust the policies by changing coefficients in the cost function according to different mission' needs. As a result, RL is more adaptive to various scenarios and is more suitable to be a base for building the mobile edge computing platform.

Motivated by the reasons mentioned above, we propose a platform to advance research in UAV-mounted mobile edge computing by building a unified framework with path planning algorithms based on RL.
The main contributions of this paper are summarized as follows:

\begin{itemize}
	\item
	First, we provide a novel framework in which UAV-mounted mobile edge computing and path planning are combined based on RL by considering the geometric distance, risk, and terminal users' demand in a single cost matrix.
	\item
	Second, we investigate multi-UAV collaboration in the mobile edge computing scenario. Geometric and terminal user's information are shared among UAVs, thus ensuring cost-saving and obstacle avoidance.
	\item
	Third, we introduce an efficient way of depicting the terminal users' demand for achieving a higher $QoS$. Compared with the traditional linear demand function, the sigmoid-like function enables better task allocation.
	\item
	Fourth, we perform extensive experiments to test the proposed platform and evaluate different coefficients in the cost function. Results have shown the effectiveness and feasibility of our approach.
\end{itemize}

The rest of this paper is organized as follows. First, related work is presented in Section \uppercase\expandafter{\romannumeral2}. Then in Section \uppercase\expandafter{\romannumeral3}, we provide a detailed description of the UAV-mounted mobile edge computing platform. Finally, the effectiveness and feasibility of the proposed platform are validated by simulations in Section \uppercase\expandafter{\romannumeral4}, and conclusions are presented in Section \uppercase\expandafter{\romannumeral5}.

\section{Related Work}
We conduct our literature review to cover mobile edge computing, path planning, and their combination.

\subsection{Mobile edge computing}

Mobile edge computing has attracted increased interest and is becoming one of the hottest topics in edge computing. For example, \cite{beck2014mobile} develops a taxonomy of mobile edge computing applications and use cases. In addition, there some representative reviews: \cite{abbas2017mobile} presents related concepts and technologies, architectures, advantages, and typical scenarios of mobile edge computing,  \cite{mach2017mobile} explains architecture and computation offloading, and \cite{mao2017survey} investigates mobile edge computing communications. These works demonstrate the potential of this field.

While the concept of mobile edge computing has been seen in the literature before, it remains an open problem. According to \cite{mach2017mobile}, the distribution and management of mobile edge computing resources is a key requirement for ensuring the $QoS$ of terminal users. When servers are moving dynamically, the system benefits from flexibility but at the same time becomes more complicated, which deteriorates the incapability of most arrangement methods. 

\subsection{Path planning}

Kim et al. \cite{kim2014coordinated} introduce a path planning algorithm for dynamic environments where small UAVs are used as relay nodes in a network of naval vessels. The motion estimates of the vessels and the states of the UAVs are taken as input to generalize the strategy. Instead of optimizing a centralized system, the approach exploits a fully decentralized non-linear model predictive control concept. To emphasize the cooperation between UAVs, Zhang et al. \cite{zhang2014cooperative} propose the Cooperative and Geometric Learning Algorithm (CGLA)  designed for path planning based on the cooperation of multiple UAVs. CGLA introduces a weight matrix based on geometric distance and integral risk information to guide the movement of UAVs. The weight matrix can be efficiently calculated and updated, making the system lighter than systems based on methods such as neural networks and guarantees real-time path planning. We note that there is a relatively low requirement for computing power in CGLA reinforcement learning, and this makes it a suitable approach for UAV-mounted mobile edge computing.

\subsection{Combination of mobile edge computing and path planning}

Previous works introduce successive convex approximation (SCA) to combine mobile edge computing and path planning. Jeong et al. \cite{jeong2017mobile} leverage SCA strategies to calculate the path of UAVs under latency and UAV's energy budget constraints.  \cite{cao2018mobile} investigates a scenario where the UAV offloads its computation tasks to multiple ground stations along its trajectory. The authors exploit alternating optimization and SCA techniques to design the UAVs trajectory to minimize the mission completion time. However, these works do not involve a cooperative mechanism between UAVs. In addition, these approaches have a limitation when the environment is unknown in advance. The learning-based algorithms also attract the attention of path planning in UAV-mounted mobile edge computing. For example,  \cite{cheng2019space} investigates a joint task scheduling and resource allocation approach in a space-air-ground integrated network based on policy gradients and actor-critic methods.
 \cite{yin2019intelligent} applies a deterministic policy gradient algorithm to maximize the uplink sum rate in the UAV-aided cellular network with multiple ground users.

Inspired by these works, we provide an organic combination of mobile edge computing and path planning by building an open-source platform for the UAV-mounted mobile edge computing networks. Simulation results have shown the feasibility and flexibility of our platform.

\section{Multi-UAV Mobile Edge Computing and Path Planning based on Reinforcement Learning}

In reinforcement learning, the agent's goal is to find an optimal strategy under each scenario to obtain the maximum expected reward. In our platform, the mobile network processor UAV is the agent who constantly learns from the environment. 
At each time slot, a UAV chooses a planning strategy to achieve the best possible reward according to its surroundings. After the UAV moves, the surrounding changes and positive or negative feedback is provided to the UAV in the form of a reward matrix $A$ given the factors of risk, geometric distance, and terminal users’ demand. The UAV then learns from the surroundings through a stochastic iterative cost matrix $G$ generated from $A$ and selects a strategy---a path toward the target. $G$ can be considered as the memory of each agent, which is intensified and ‘trained’ through each episode of the planning process.

\subsection{Environment modeling}

This paper considers the UAV's collision avoidance and terminal users' demand on the same platform. The surroundings contain two basic elements of obstacles and terminal users. First, obstacles vary in shape, positions, and risk levels, which include buildings, cars, or mountains in a real environment. Second, we assume the obstacles comply with a Gaussian distribution but have different variances $\sigma$ used to calculate their risk exposure probability.

For $n$ independent obstacles in the map, giving the $i$th obstacle position $O_i=(X_i,Y_i)$, the risk $r_i(x,y)$ indicates the risk from $O_i$ at the point $(x,y)$ and can be defined as

\begin{equation}
    \begin{split}
    r_i(x,y)=\frac{1}{\sqrt{2\pi}\sigma}e^{-\frac{d^2}{2\sigma}},\ d=\sqrt{(x-X_i)^2+(y-Y_i)^2},\\
    i\in\left\{1,2,\cdots,n\right\}.
    \end{split}
    \label{eq1}
\end{equation}

Considering all $n$ obstacles in the map, the overall risk to a point $(x,y)$ in the risk exposure probability matrix can be described as
\begin{equation}
    R(x,y)=1-\prod_{i=1}^{n}[1-r_i(x,y)].
    \label{eq2}
\end{equation}
The exposed risk from any point $p$ to any point $q$ on the map is the integral risk of $R(x,y)$ for any $(x,y)$ on $C$, where $C$ is the linear path from $p$ to $q$:

\begin{equation}
    \int_{(x,y)\in C}^{} R(x,y).
    \label{eq3}
\end{equation}

\begin{figure}[h]
    \centering
    \includegraphics[width=0.9\linewidth]{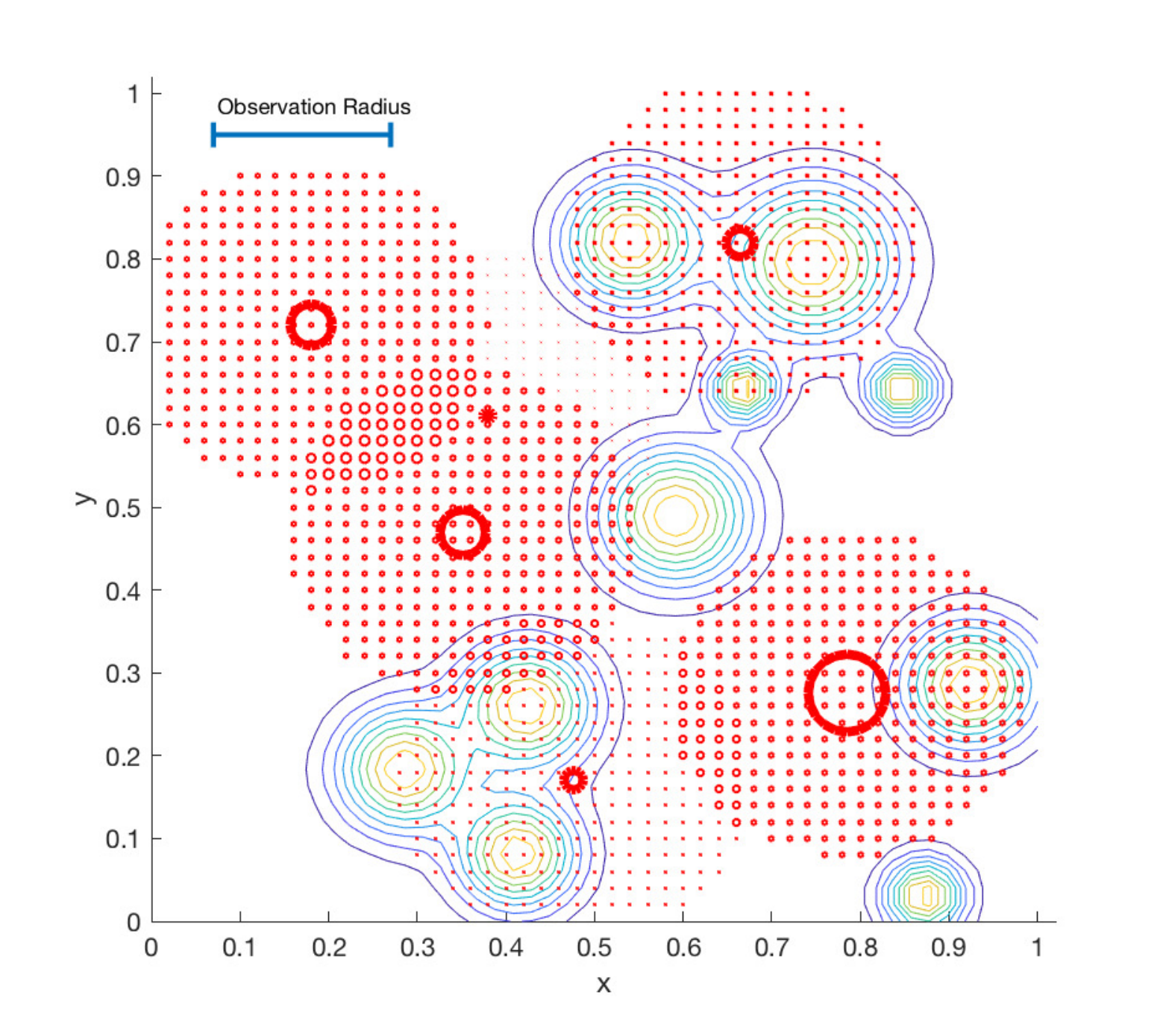}
    \caption{Obstacles’ risk and terminal users’ demand map ($\epsilon=0.2$)}
    \label{map}
\end{figure}

Second,  for serving terminal users, we assume that each terminal user has an initial demand $d^0_j$ for the UAVs to process. We also assume the demand can only be served by UAVs within a constant service radius because the UAVs have limited capability for detecting demand signals beyond a certain distance. Therefore, the service area is denoted $s(p_j,\epsilon)$, where $p_j$ is the position of $TU_j$, and $\epsilon$ is the service radius, as shown in Figure \ref{map}.

When a UAV enters the service range of $TU_j$, the service for $TU_j$ begins. The remaining demand of $TU_j$ will decrease at a constant speed $\tau$ per unit time per UAV. We can easily infer that a terminal user with a greater demand needs more time to be served, and the longer a UAV remains in the service range of $TU_j$, the more service can be provided for $TU_j$. $d_j$ changes over time $t_k$ time served by $UAV_k$ as

\begin{equation}
    d^{l+1}_j = d^l_j - \tau t_k.
    \label{eq4}
\end{equation}

The correlation between the terminal users' demand and UAVs' demand detection should have a non-linear relationship to improve system performance. With reference to \cite{liu2020path} and \cite{lee2005non}, a sigmoid-like function can help enhance strong signals and abate weak signals. Thus, we adopt a sigmoid-like demand detection function $U(d_j)\in(0,1]$ to describe the correlation between real demand and detected demand as
\begin{equation}
    U(d_j)=1-{\rm exp } \left [ -\frac{(d_j)^\eta }{d_j+\beta}  \right ],
    \label{eq5}
\end{equation}
where $\eta$ and $\beta$ are controlling variables.  

\begin{figure}[h]
    \centering
    \subfigure[Single sigmoid function]{
    \includegraphics[width=0.9\linewidth]{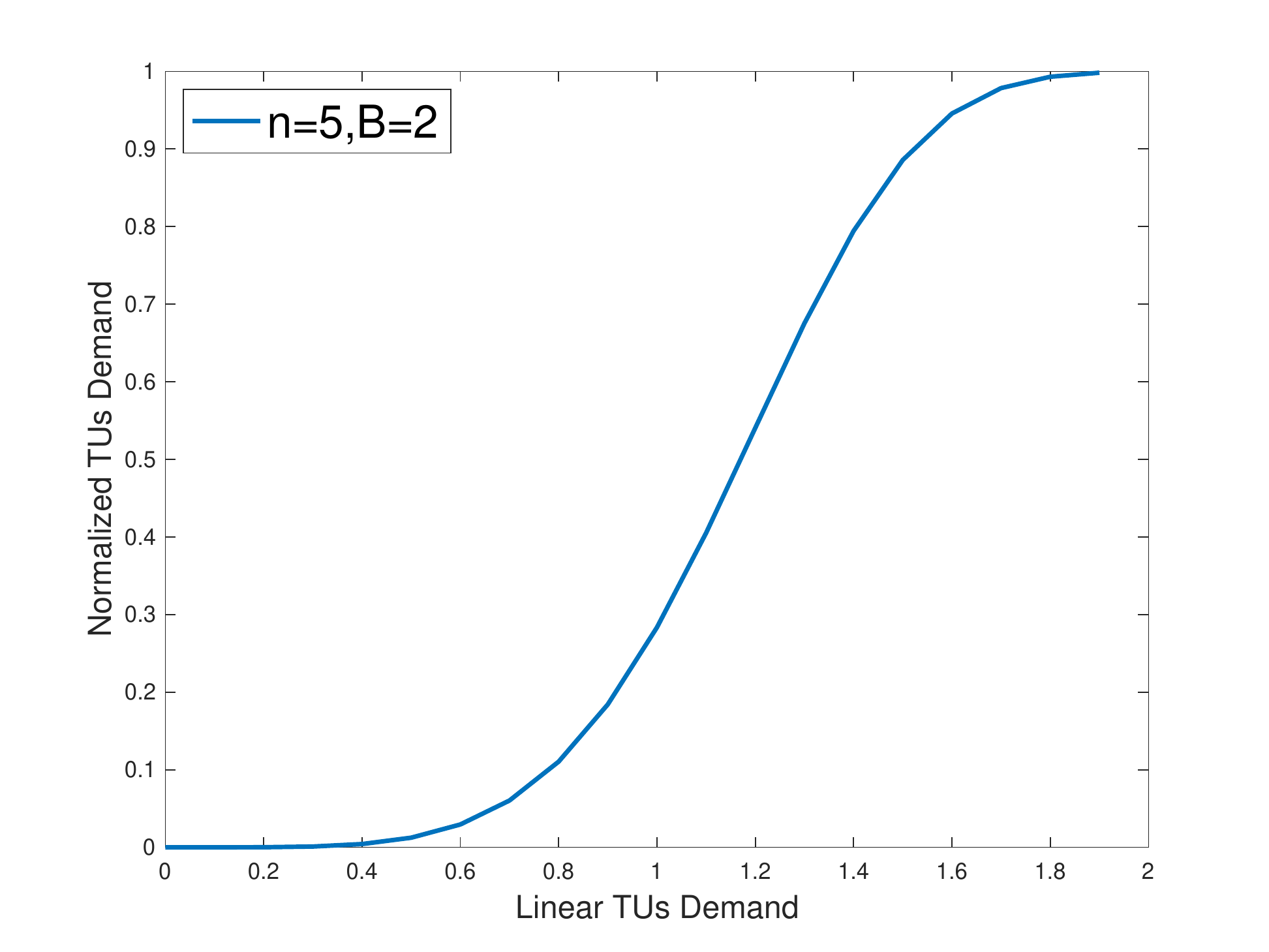}
    \label{sig1}}
    \subfigure[Sigmoid functions with different parameters]{
    \includegraphics[width=0.9\linewidth]{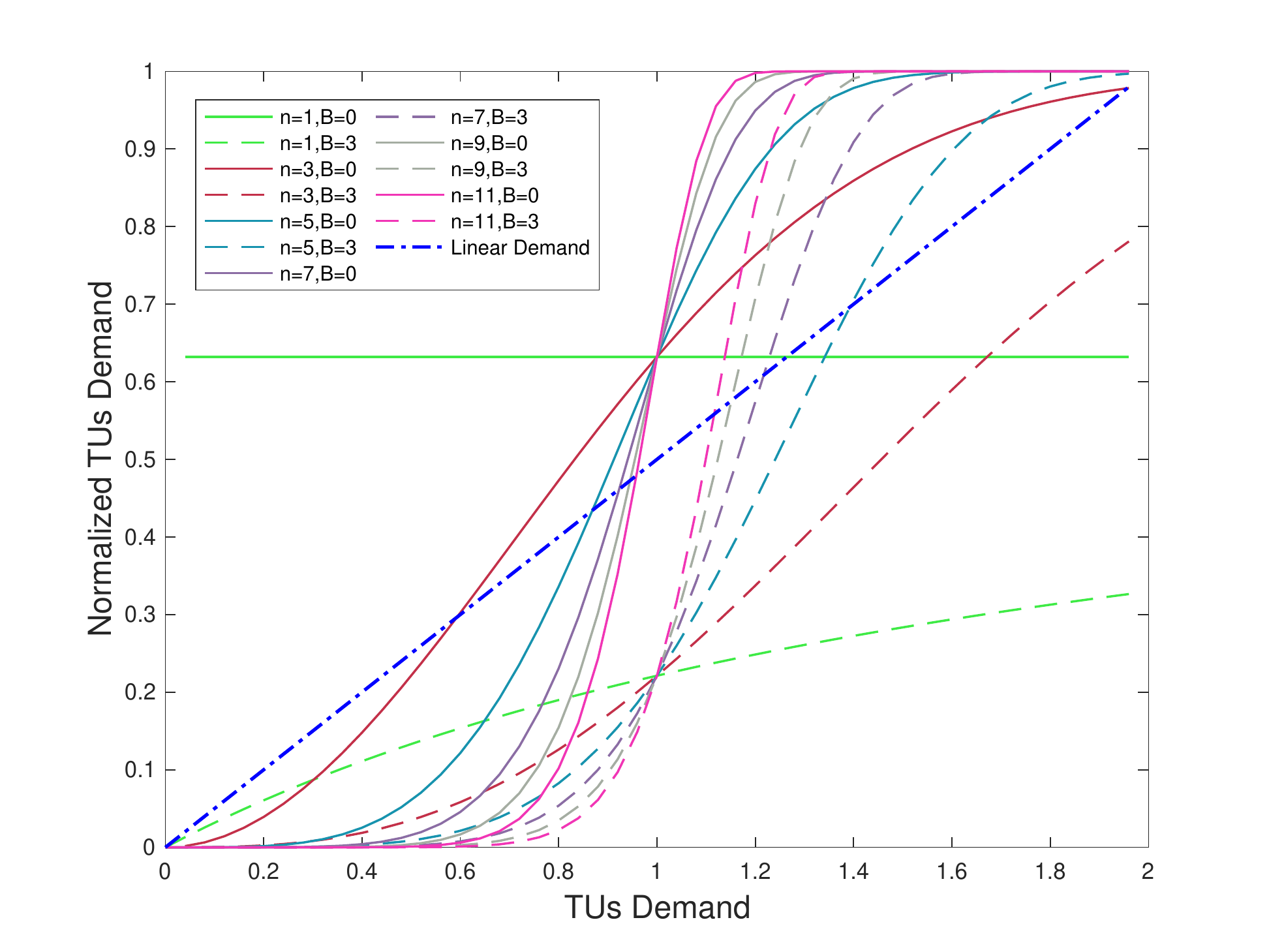}
    \label{sig2}}
    \caption{Examples of normalized sigmoid demand functions}
    \label{sigmoid}
\end{figure}

From \ref{sig1}, $U(d_j)$ first increases steeply as the demand rises and becomes steady when the demand is sufficiently great. Thus,  \eqref{eq5} can encourage a UAV to focus on terminal users with greater unserved demand and prevent it from serving any terminal user over a long time, thus improving $QoS$.

Generally, a Sigmoid-like function is an increasing function with an inflection point $x_0$, which follows $\frac{d^2f(x)}{dx^2}>0$ when $x<x_0$ and $\frac{d^2f(x)}{dx^2}<0$ when $x>x_0$ \cite{lee2005non}. Functions with such form satisfy the following properties:

\begin{property}
     For any $x>0$ in $U(x)$, the function is valid  only when 
    \begin{center}
         $\eta\in(1,\infty)$,
    \end{center}
    \begin{center}
        $\beta\in(0,\infty)$
    \end{center}
    \label{co1}
\end{property}

The proof is given in Proof \ref{proof1} in the appendix.

\begin{property}
    $\eta$ controls the slope and centrality of the curve, pivots through an inflection point $(1,1-e^{-\frac{1}{1+\beta}})$. $\beta$ controls the horizontal movement of the curve. As a result, the intersection point can be moved vertically by changing $\beta$.
    \label{co2}
\end{property} 

The proof is given in Proof \ref{proof2} in the appendix.

As shown in Figure \ref{sig2}, $U(x)$ has an inflection point when $x = 1$. The curve becomes steeper when $\eta$ increases and moves vertically downward when $\beta$ increases. Therefore, $\eta$ and $\beta$ are constant variables that affect $QoS$. The evidence will be provided in part IV.

To simplify the model, we assume the demands are linearly accumulated. For a UAV at a point $p$ in the map, with a detection range $\epsilon$, the detected demand is the linear accumulation of terminal users' demand within circle area $s(p,\epsilon)$:
\begin{equation}
    \sum_{j\in s(p,\epsilon)}^{} U(d_j).
    \label{eq6}
\end{equation}

\subsection{The Reward matrix}

A reward matrix is introduced for UAVs to learn and adapt to find the optimal path.
The reward matrix is designed to measure the reward or punishment from any point towards any other points on the map given the factors of risk, geometric distance, and terminal users’ demand. 

In our platform, the map is represented as a lattice of $N\times N$, and the reward $A_{p_i,p_r}$ between any point $p_i$ and $p_r$ in the map is defined as

\begin{equation}
    A_{p_i,p_r} = d_{p_i,p_r} + K \int_{C}^{} R(x,y)ds + \frac{M}{1+\sum_{j\in s(p_i,\epsilon)}U(d_j)},
    \label{eq7}
\end{equation}

where $d_{p_i,p_r}$ is the geometric distance between $p_i$ and $p_r$. The second term in the equation denotes the risk detected from $p_i$ to $p_r$ or vice versa, which means the larger the risk detected, the larger the cost or punishment. The last term in the equation is the overall demand detected at $p_i$, as $p_i$ is the current position of UAV. The larger the demand detected, the smaller the punishment, or the larger the reward. 

For each point $p_r, r\in\left\{1,2,\cdots,N^2\right\}$ in the map, a reward matrix $A_{p_r}$ consists of points $A_{p_i,p_r},  i\in\left\{1,2,\cdots,N^2\right\}$ is generated with regard to all points $p_i$ on the map. $K$ and $M$ reflect the tolerance of risk and priority of service, which affects the strategy for path planning. When applied in real situations, $K$ and $M$ can be adjusted according to the  mission's requirements. For instance, if $K$ is set to a relatively high value, the UAVs will tend to stay away from obstacles even though this leads to a longer path length.

\begin{figure}[h]
    \centering
    \subfigure[$UAV~pos=(0.15,0.4)$]{
    \includegraphics[width=0.46\linewidth]{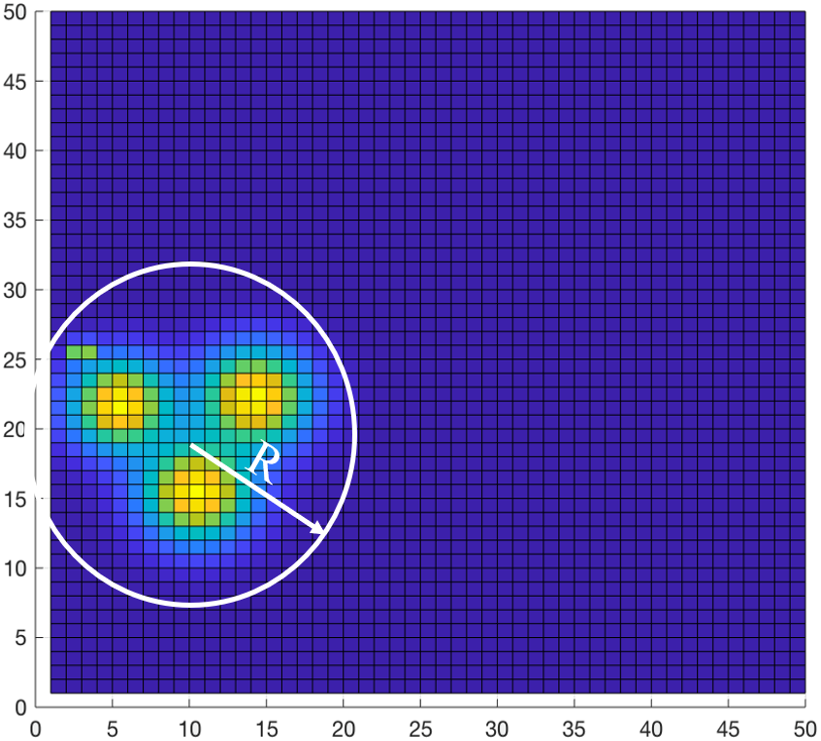}
    \label{E1}}
    \subfigure[$UAV~pos=(0.75,0.6)$]{
    \includegraphics[width=0.46\linewidth]{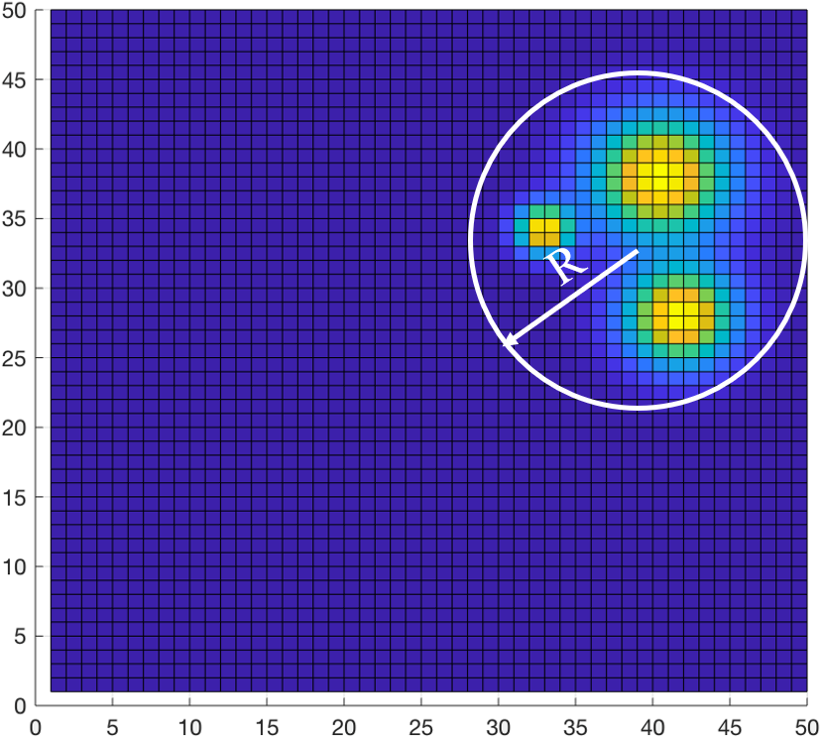}
    \label{E2}}
    \caption{Observed risk distribution ($R=0.2$)}
    \label{observed_E}
\end{figure}

An obstacle observation radius is introduced for each UAV to fit real situations. When an obstacle enters the observed area of a UAV, the UAV detects the obstacle and obtains risk information. Only the observed obstacles will count as a risk when calculating the weight matrix. The highlighted color in Figure \ref{observed_E} shows the observed risk calculated by \eqref{eq1} and \eqref{eq2} for a UAV in position $P$ with an observation radius $R$.

\begin{figure}[h]
    \centering
    \subfigure[$UAV~pos=(0.15,0.4)$]{
    \includegraphics[width=0.46\linewidth]{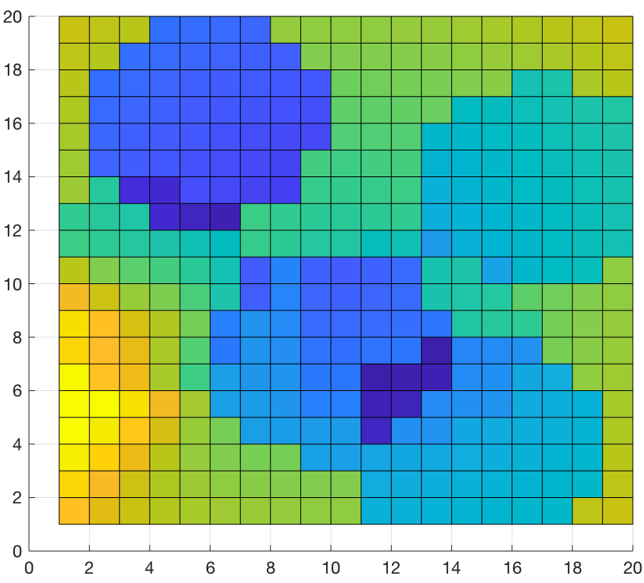}
    \label{A1}}
    \subfigure[$UAV~pos=(0.75,0.6)$]{
    \includegraphics[width=0.46\linewidth]{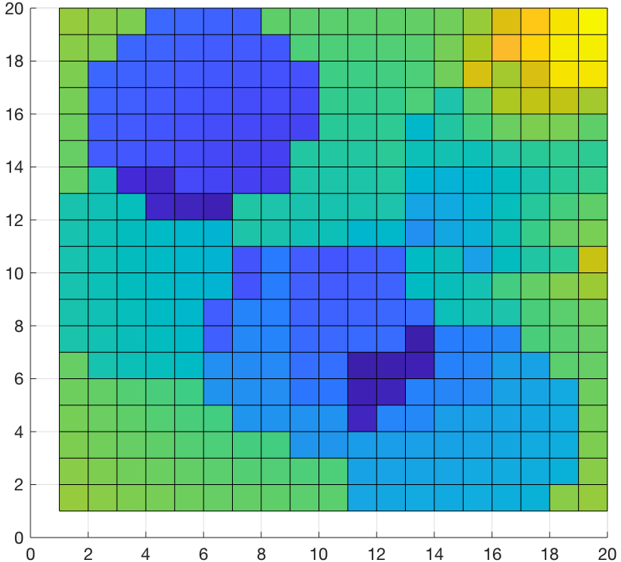}
    \label{A2}}
    \caption{Weight matrix on point $(0.5,0.5)$}
    \label{observed_A}
\end{figure}

Figure \ref{observed_A} $(a),(b)$ shows the reward matrix on point $[0.5,0.5]$ based on each UAV's observed risk distribution in Figure \ref{observed_E} and the overall terminal user's demand. According to \eqref{eq7}, the terminal user's demand decreases the cost, and the observed risk adds the cost. In Figure \ref{observed_A}, we can infer that places with more terminal user demand have lower values (dark color), and those with more obstacles have higher values (light color).

\subsection{Cost matrix}
\label{cost matrix}
During the path planning process, a cost matrix $G$ is introduced for a UAV to obtain a preliminary optimal path to the destination. 
The generation of the cost matrix is obtained using an iterative process. After several iterations, the cost matrix will converge\cite{zhang2014cooperative} and remain steady. The update mechanism of the cost matrix in a map with $N \times N$ computing nodes is described as:

\begin{enumerate}
    \item 
    \texttt{Initialize G}: initialize the cost matrix $G^0$. Assign the value $0$ to the target point and the value $\infty$ to all other points.
    \item 
    Update the cost matrix $G$. Randomly choose a position $p_r$ from the map. For each point $p_i$ in the map, update the point value in $G$ by comparing the current value with the revised value considering the reward matrix:
    \begin{equation}
    \begin{split}
        G_{p_i}^{k+1}=min\left \{G_{p_i}^{k}, A_{p_i,p_r}+G_{p_r}^k  \right \},\\ i,r\in \left \{ 1,2,\cdots,N^2 \right \}.
    \end{split}
    \label{eq8}
    \end{equation}
    \item 
    Repeat step 2) until reaching the maximum number of iterations.
\end{enumerate}

After $G$ is produced, an ordinal sequence of points is generated as a preliminary $Path$ for the UAV to follow. Each UAV has its own $Path$.
Points with the lowest cost in $G$ are constantly added. The generation of $Path$ is described as follows:
\begin{enumerate}
    \item 
    Initialize $Path$ as an empty list.
    \item 
    Add $p_i$ with the lowest value in $G$ to $Path$, then assign $G_{p_i}$ to $\infty$.
    \item
    Repeat step 2) until reaching the target point or reaching a maximum length.
\end{enumerate}

We note that the elements in $Path$ are in ascending order of cost. The process of calculating $G$ and generating $Path$ together form the function \texttt{Planning} in Algorithm \ref{algo1}.

\subsection{UAVs Movement}

Each UAV is considered an agent in the RL process. They are therefore assigned a memory $D_i$ and a cost matrix $G^i$. $D_i$ stores map information and serves as the "eyes" and memory of the agent, and $G^i$ serves as the "brain" of the agent. The agent produces a learned result $Path_i$ to complete each episode.
All UAVs in the system move in sequence to realize information sharing. This can result in a slight time delay in real situations. When one UAV moves, the other UAVs would be treated as obstacles.
Algorithm \ref{algo1} describes the process. For the $i^{th}$ UAV ($UAV_i$), after moving one step according to $Path_i$, \texttt{ScanEnv} is performed. In \texttt{ScanEnv}, $UAV_i$ scans the circle area $s(pos_i, R)$ where $R$ is the observation radius. This is used to decide whether further planning is needed. If new obstacles, including other UAVs, are observed, $ObstacleFound$ will be set to $True$, and memory $D_i$ will be updated. Then, the weight matrix and $G^i$ will be altered, and $Path_i$ will be recalculated. If the surroundings remain unchanged, $UAV_i$ continues to \texttt{Move} according to $Path_i$. In \texttt{Move}, $UAV_i$ moves a distance of $StepLength$ along the direction of the vector starting from $pos_i$ and ending at $Path_i[1]$, if the distance between $pos_i$ and $Path_i[1]$ is smaller than $StepLength$, the UAV will move directly to $Path_i[1]$. \texttt{Move} returns the new position of $UAV_i$. At the end of one loop, the remaining demands of all TUs within the service area of $UAV_i$ are updated according to \eqref{eq4}.

 \begin{algorithm}[h]
    \caption{UAV Movement Algorithm}
    \begin{algorithmic}[1]
        \FOR{$i$ in $UAVnum$}
            \STATE \texttt{Initialize G($i$)}
            \STATE $Path_i \gets$ \texttt{Planning()}
        \ENDFOR
        \FOR{$i$ in $UAVnum$}
            \IF{$pos_i==TargetPoint$}
                \STATE \texttt{Stopmovement($i$)}
            \ELSE
                \STATE \textit{// Remove outdated information from $D_i$ because $pos_j$ has changed in last loop}
                \FOR{$j$ in $UAVnum$ and $j!=i$}
                    \STATE delete $pos_j$ from memory $D_i$
                \ENDFOR
                \STATE $ObstacleFound \gets$ \texttt{ScanEnv($pos_i, R$)}
                \IF{$ObstacleFound$}
                    \STATE $Path_i \gets$ \texttt{Planning()}
                \ENDIF
                \IF{$pos_i == Path_i[1]$}
                    \STATE $Path_i \gets Path_i[2...end]$
                \ENDIF
                \STATE $pos_i \gets$ \texttt{Move($StepLength, Path_i[1]$)}
                \FOR{$TU_j$ within $s(pos_i,\epsilon)$}
                    \STATE $d_j \gets d_j - \tau$
                \ENDFOR
            \ENDIF
        \ENDFOR
    \end{algorithmic}
    \label{algo1}
\end{algorithm}

\section{Simulation and Discussion}

This section first illustrates the UAV dynamic planning process and discusses the influence of the parameters $K$ and $M$ in \eqref{eq7} on the planning results. Second, we demonstrate the efficiency of the sigmoid demand function \eqref{eq4} by comparing it with a linear demand function. Finally, we compare our algorithm with a commonly used algorithm A*, which shows that our planning algorithms achieve a much better result in terms of $QoS$.

To demonstrate our algorithm, we have made several simplifying assumptions:
\begin{itemize}
    \item The site is abstracted to grids so that objects, including obstacles, terminal users, and UAVs, are aligned with the grid.
    
    \item All object information is stored in databases for UAVs to 'scan,' while in real situations, the scanning process could be realized by perception algorithms based on sensors like cameras or radars.
    
    \item The variation in speed is realized through adjusting $StepLength$. In our proposed algorithm, each UAV takes 1 unit of time to move one $StepLength$. If the distance between a UAV’s current position and its planned position is smaller than  $StepLength$, the UAV will move to the planned position instead of moving one $StepLength$. In practical situations, the calculation distance in a map could be set to smaller than one $StepLength$, which would require speed variations.  However, we are assuming a 2D planning scenario. The altitude of the UAV in a 3D environment is not considered.
\end{itemize}

\subsection{Path Planning for Multiple UAVs}

In this simulation, we set $K = 20$, $M = 1$, $\eta=2$, $\beta=8$, $\epsilon, R=0.2$. $K, M$ is decided by users of the platform with different needs, while other parameters are determined by practical conditions and the UAVs' capabilities. Ten obstacles are given random positions with random variances $\sigma_{i}>0,\;i=0,1,2,\cdots,10$. Six terminal users are assigned random demands $d_j\in [0,10],\;j=0,1,2,\cdots,6$. In a real situation, the terminal user's demand could be a real-time variable.

\begin{figure}[ht]
    \centering
    \subfigure[Step 5]{
    \includegraphics[width=0.46\linewidth]{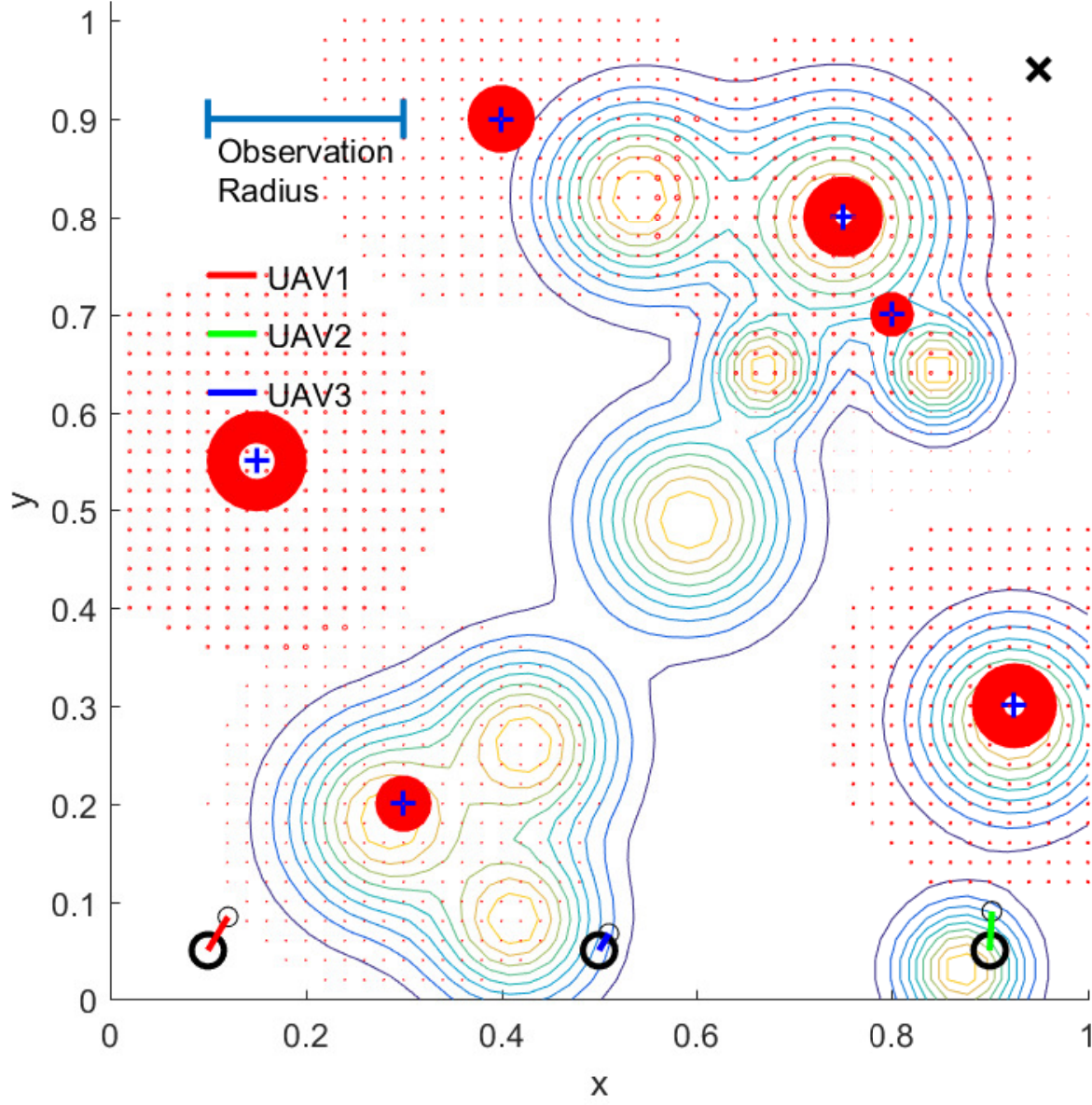}
    \label{step5}}
    \subfigure[Step 50]{
    \includegraphics[width=0.46\linewidth]{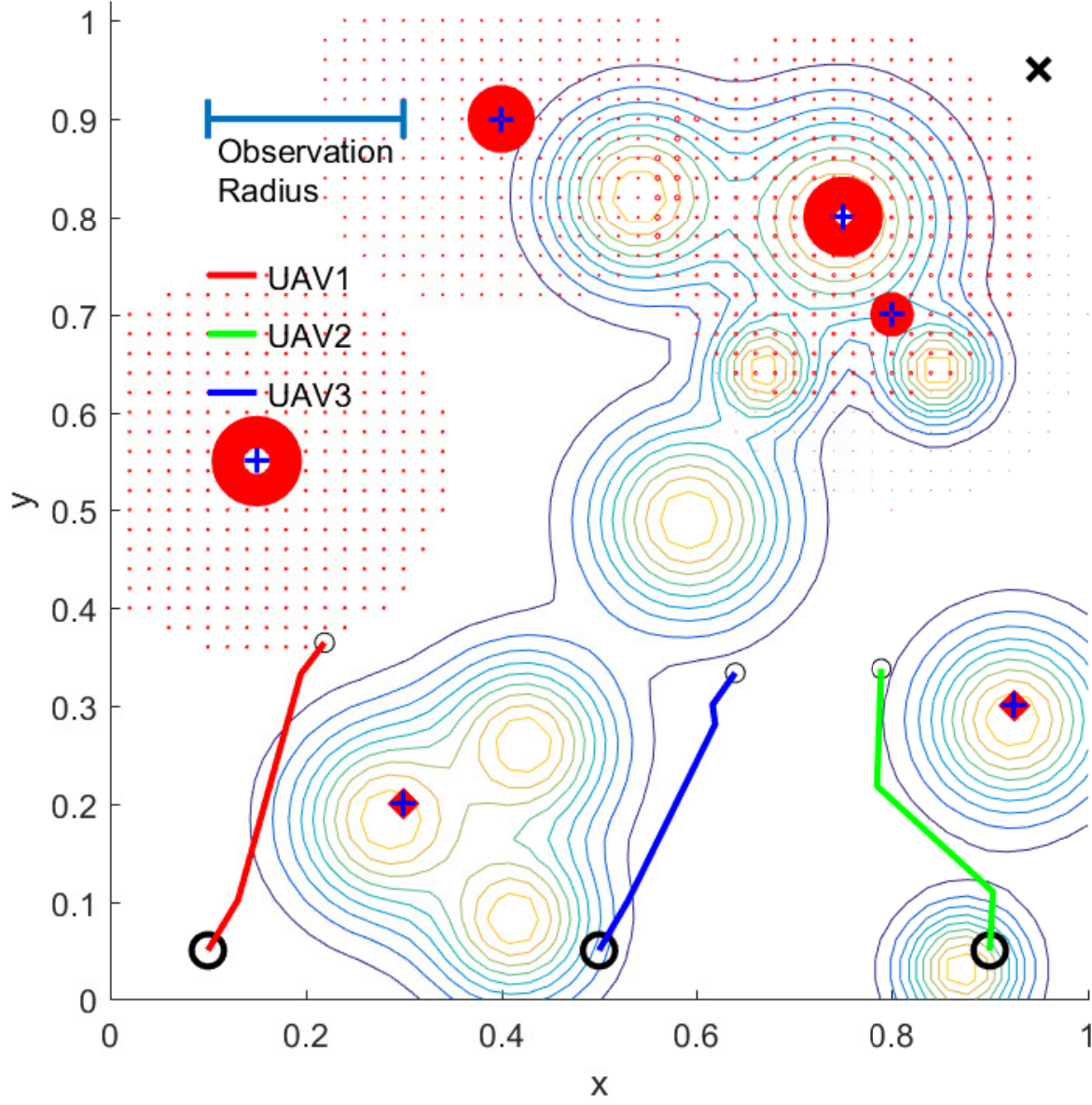}
    \label{step50}}
    \subfigure[Step 120]{
    \includegraphics[width=0.46\linewidth]{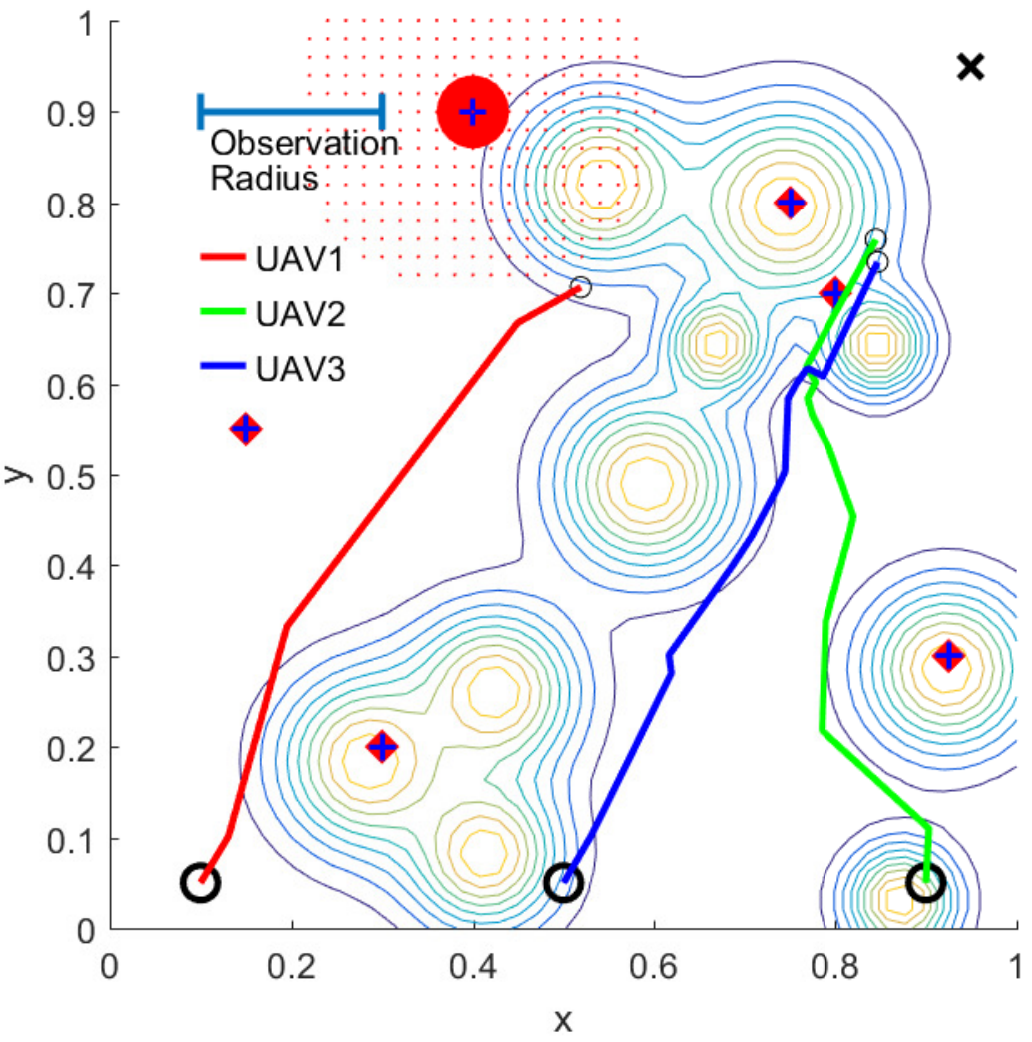}
    \label{step120}}
    \subfigure[Step 150]{
    \includegraphics[width=0.46\linewidth]{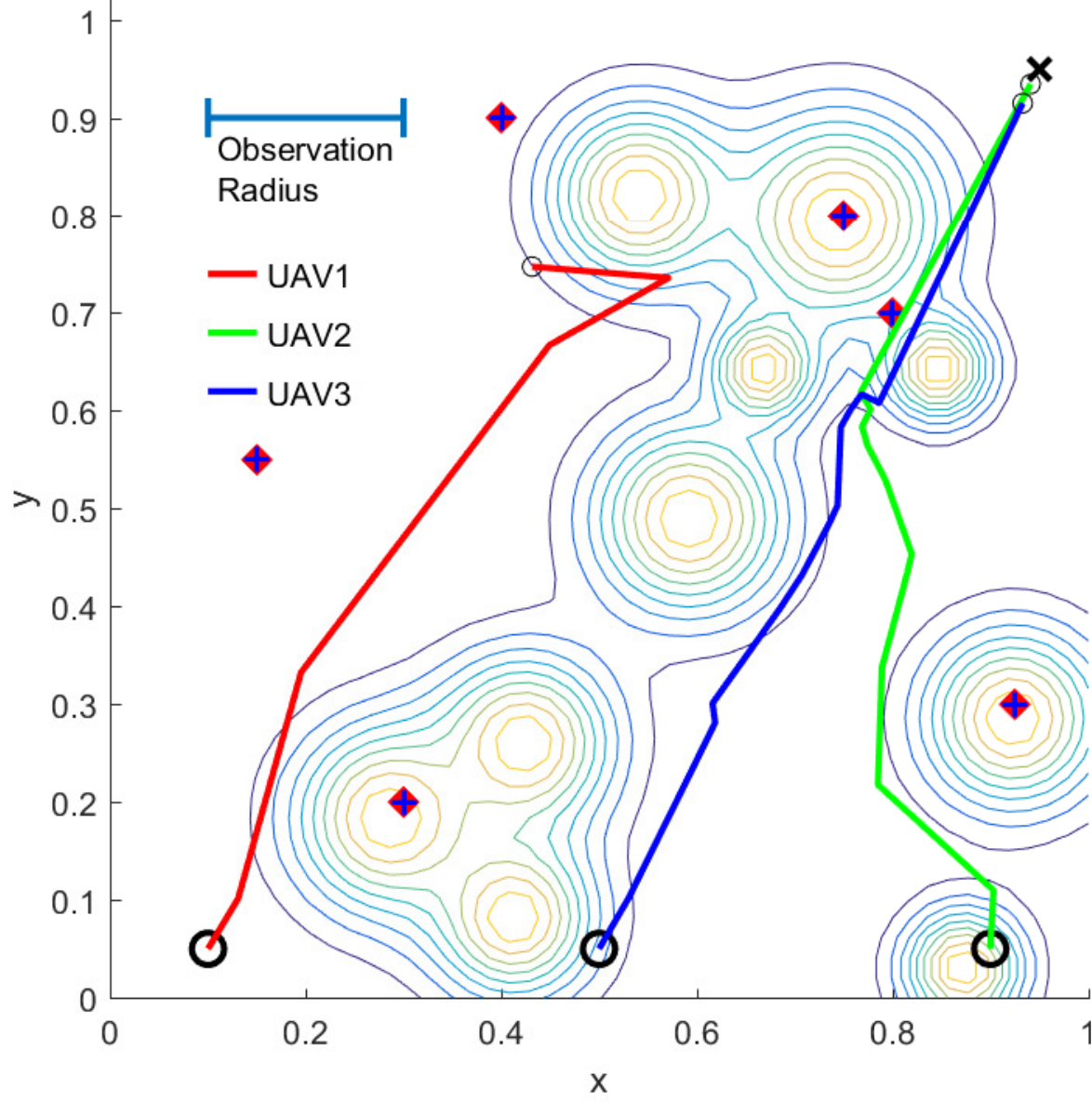}
    \label{step150}}
    \subfigure[Step 180]{
    \includegraphics[width=0.46\linewidth]{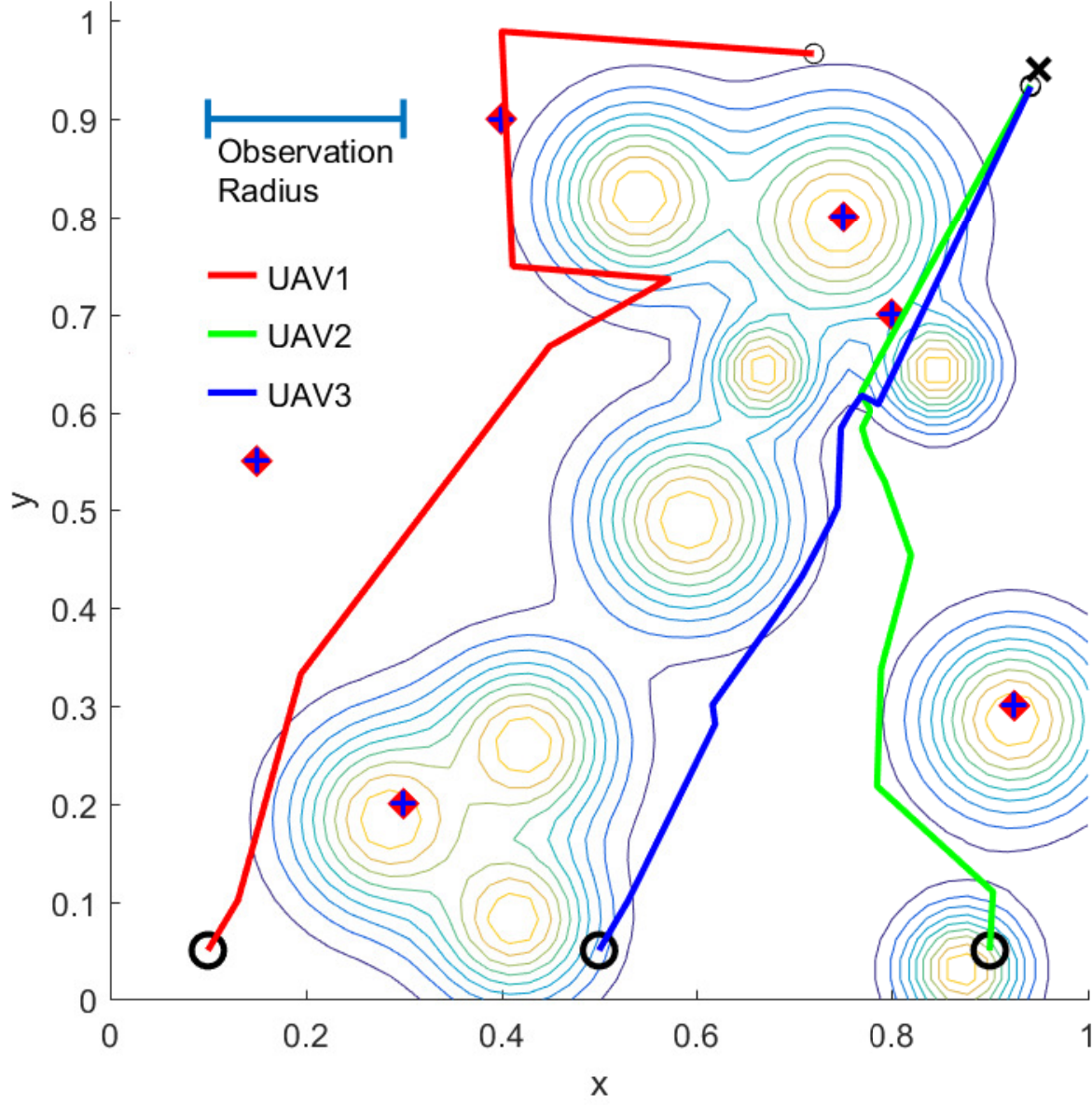}
    \label{step180}}
    \subfigure[Step 191]{
    \includegraphics[width=0.46\linewidth]{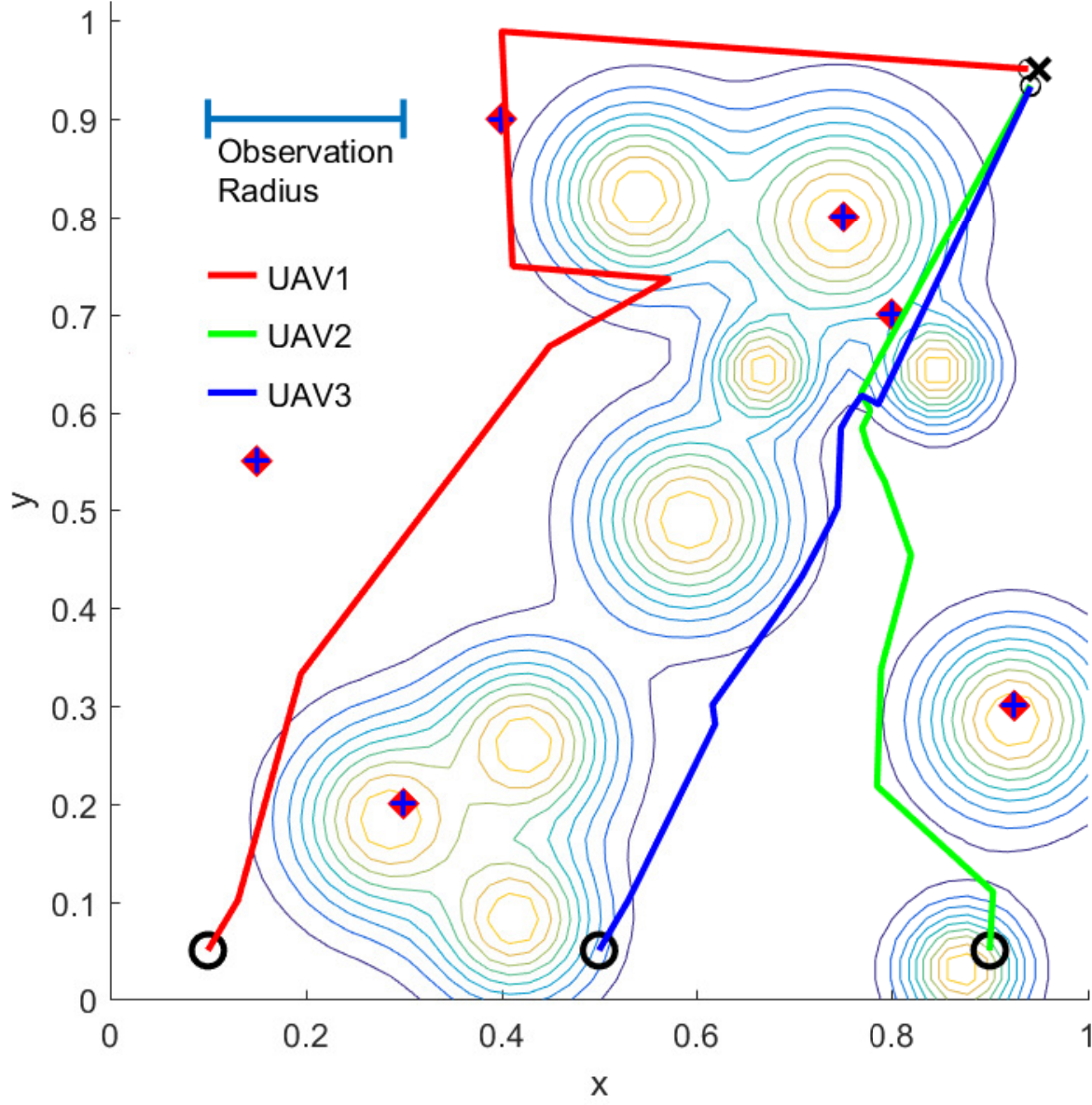}
    \label{step191}}
    \caption{Simulation of UAV-mounted mobile edge computing based on proposed path planning platform}
    \label{planning process}
\end{figure}

The planning process of the three UAVs is shown in Figure \ref{planning process}. Note that:
\begin{itemize}
    \item The point with the black cross marks the target point for all UAVs. All UAVs are tasked to deliver service to terminal users on the map and fly to the target point for each mission.
    \item The red dots show the existence and the amount of the terminal users' demand with a  service radius $\epsilon$. The red dots shrink while terminal users are being served, representing a reduction in the remaining demand. As can be seen from the overlap area of the service radius, the demands are cumulative. 
\end{itemize}

From Figure \ref{planning process}, we can infer that UAVs can choose a low-risk path to serve each terminal user in a complex environment based on our platform. Terminal users with higher demand are more attractive to UAVs.  After the demand is reduced, UAVs will change direction for other demand-high areas. Meanwhile, information sharing is effective in avoiding collisions between UAVs during planning.

\subsection{The Evaluation of $M$}

\begin{figure}[ht]
    \centering
    \subfigure[$M=0$]{
    \includegraphics[width=0.46\linewidth]{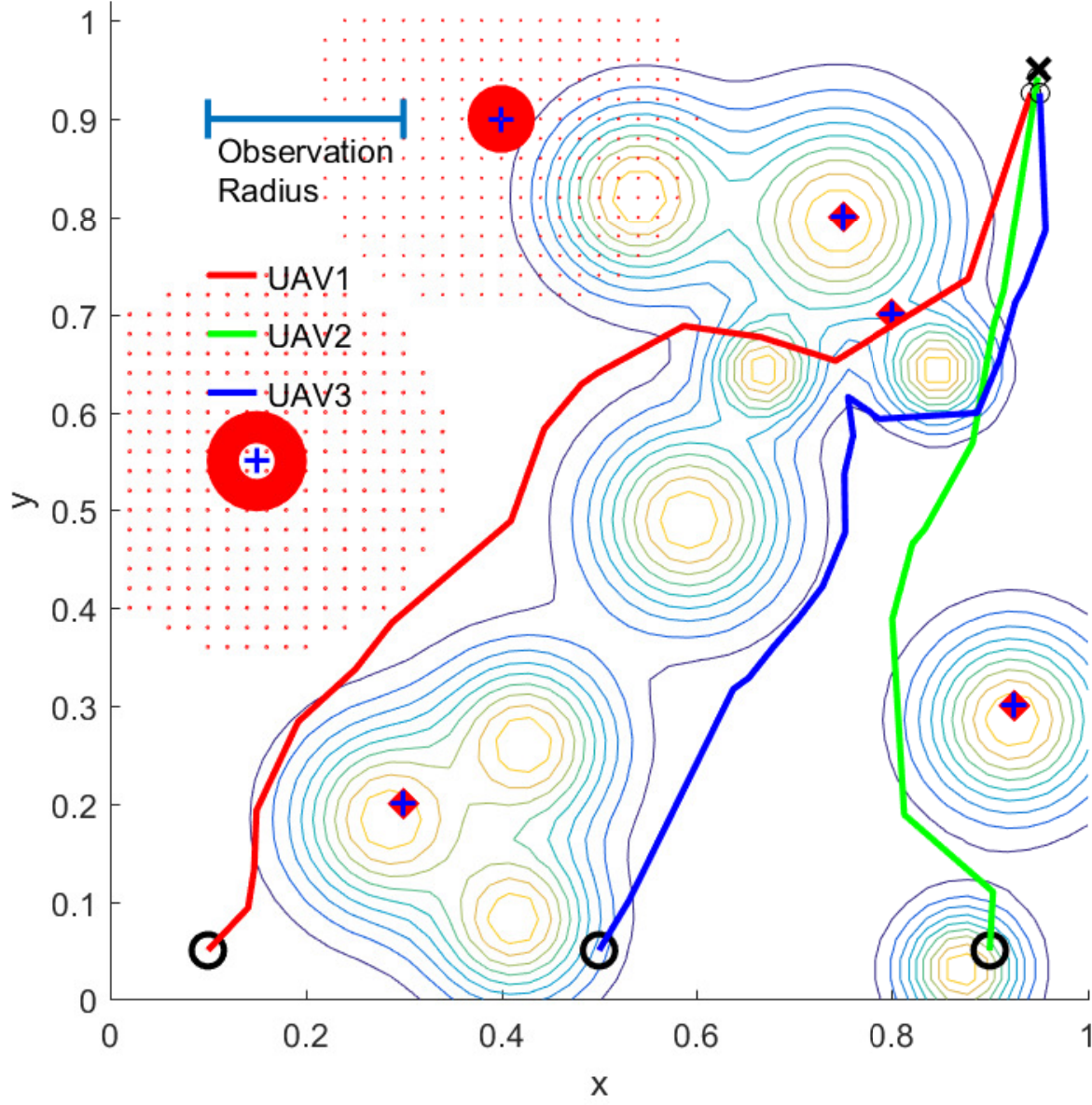}
    \label{K=5,M=0}}
    \subfigure[$M=0.01$]{
    \includegraphics[width=0.46\linewidth]{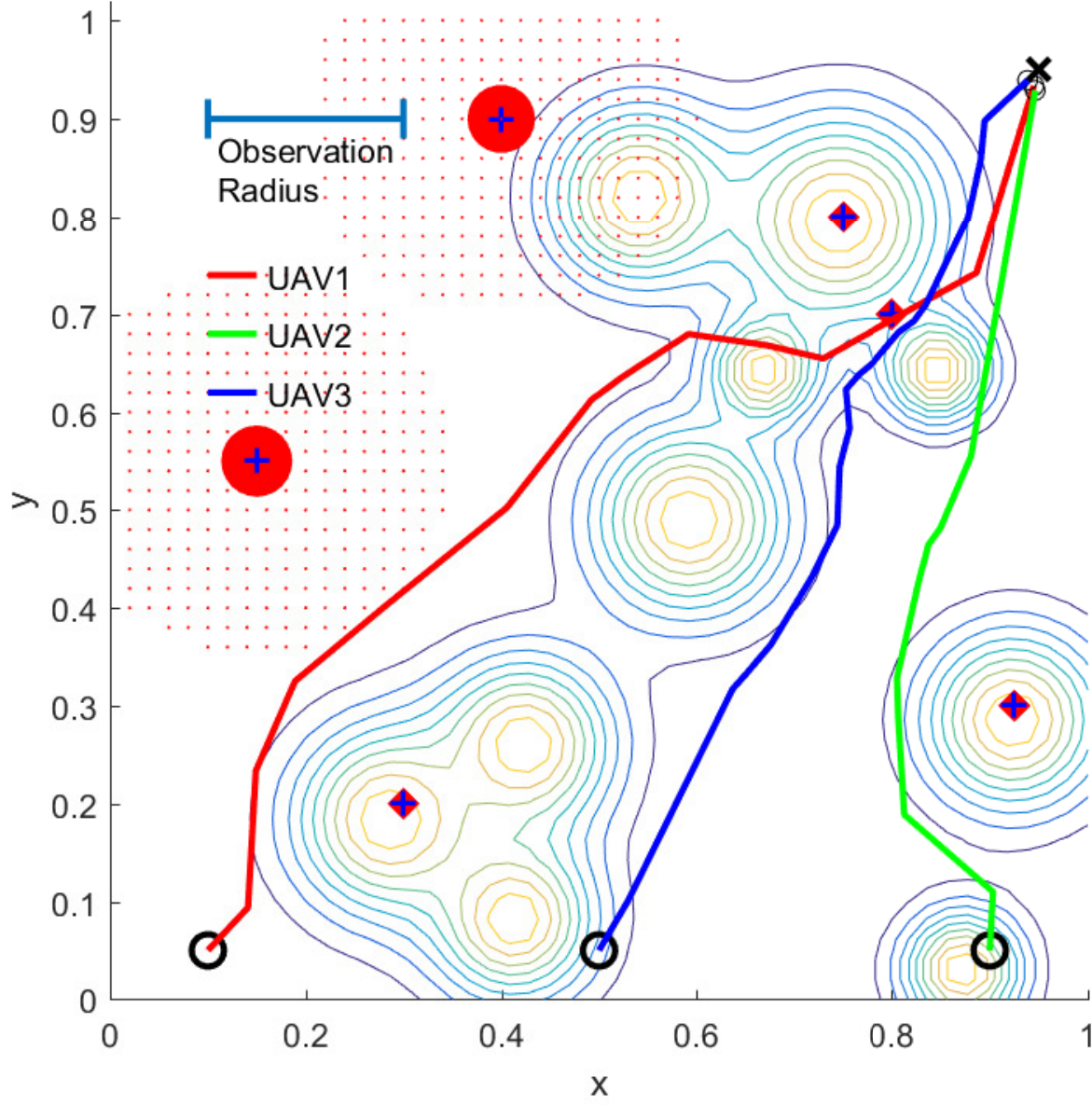}
    \label{K=5,M=0.01}}
    \subfigure[$M=0.1$]{
    \includegraphics[width=0.46\linewidth]{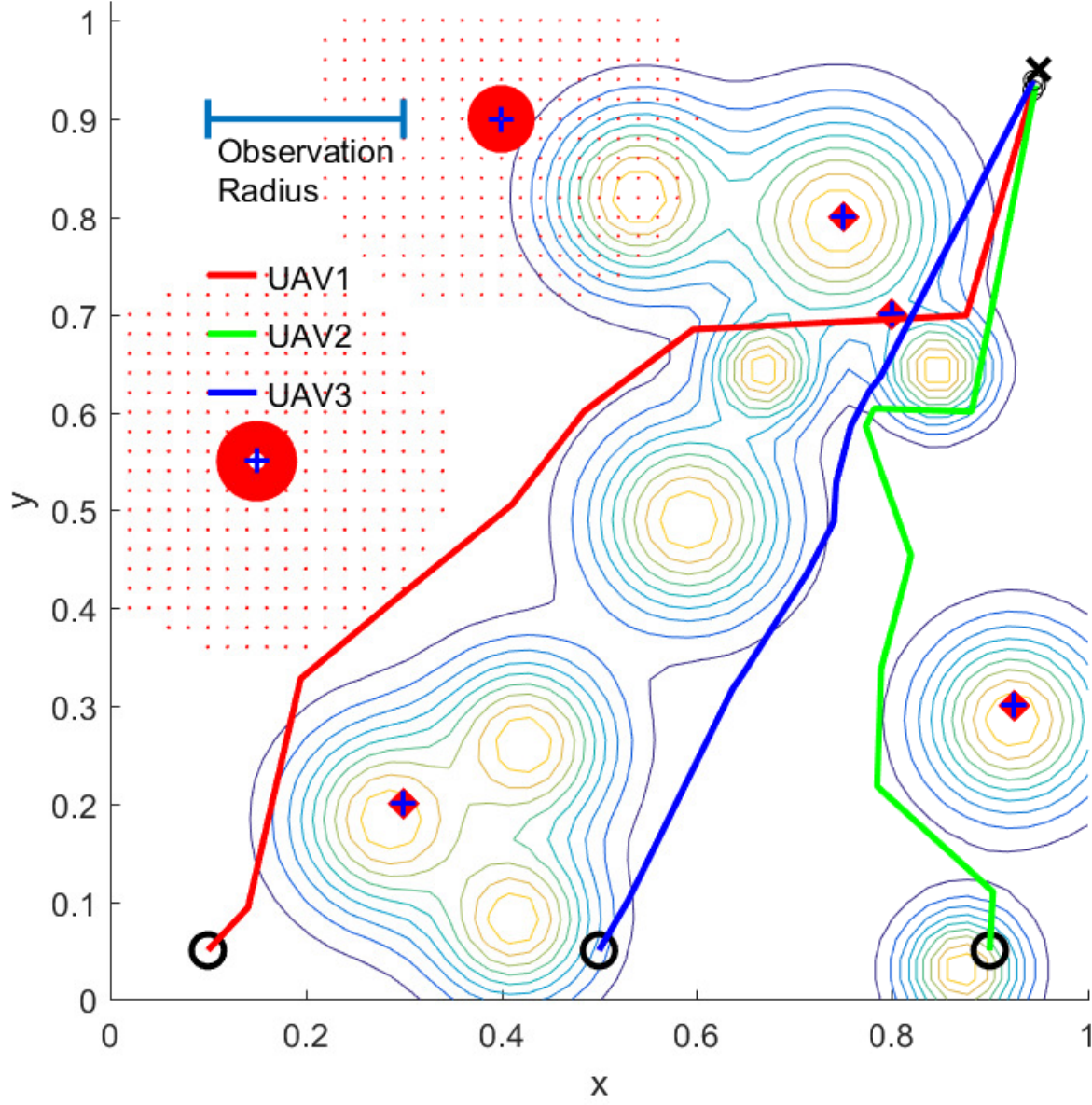}
    \label{K=5,M=0.1}}
    \subfigure[$M=0.4$]{
    \includegraphics[width=0.46\linewidth]{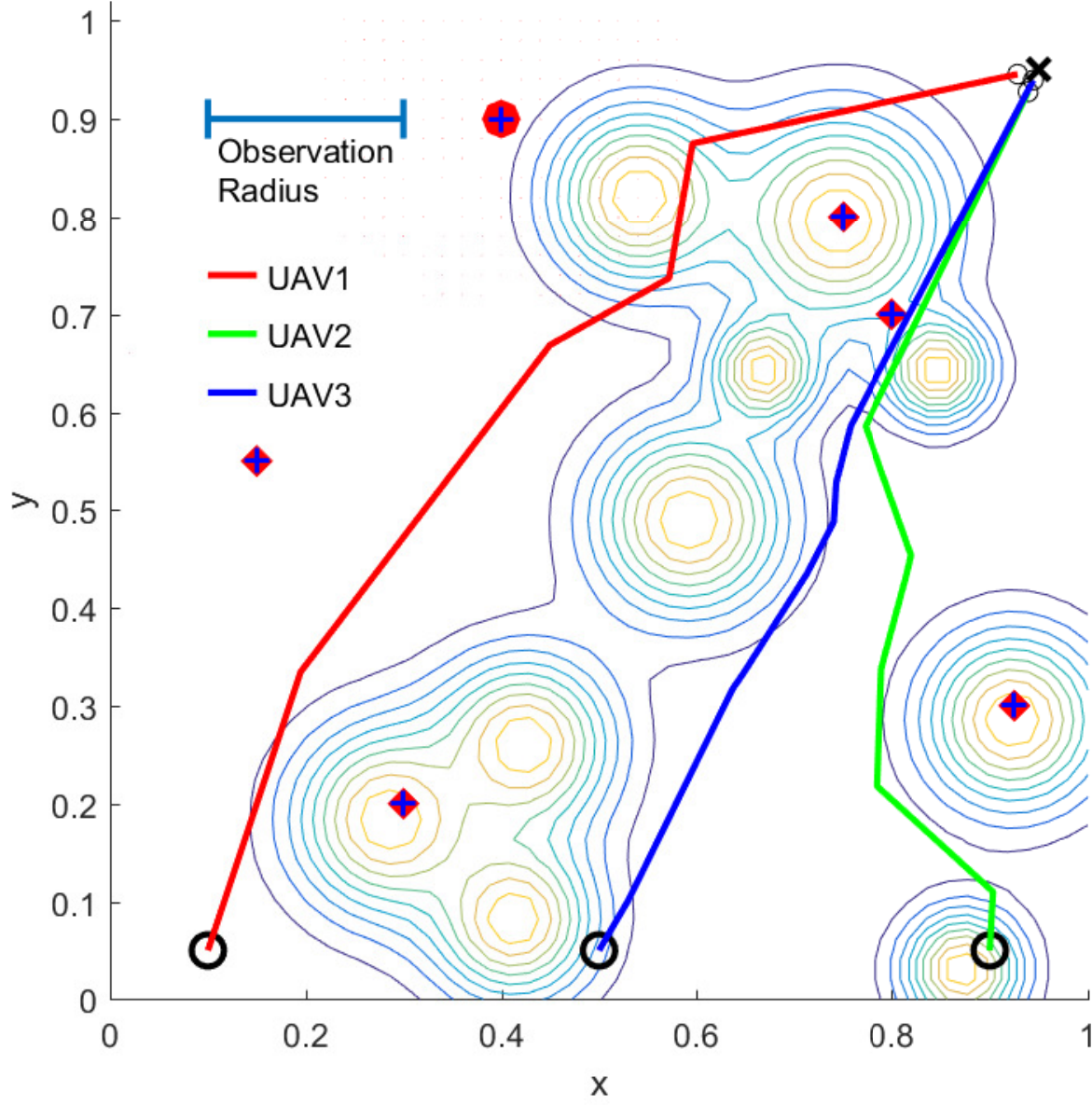}
    \label{K=5,M=0.4}}

    \caption{Planning results with different $M~(K=5)$}
    \label{4M}
\end{figure}

The parameter $M$ decides the priority of serving the terminal users and therefore controls the $QoS$. As shown in Figure \ref{4M}, given a fixed $K$, in the scenarios with larger $M$, UAVs are prone to meet more terminal users’ demands but take more risk and sacrifice path length (i.e., energy) to do so. On the contrary, UAVs in the scenarios with smaller $M$ fail to serve all terminal users but save more energy on a shorter path. 
 
The same results are also obtained by numerical simulation. To compare the service rate, we define

\begin{equation}
    {QoS}=1-\frac{\sum_{j=1}^{m}d_j}{\sum_{j=1}^{m}d_j^0}.
    \label{$QoS$}
\end{equation}

\begin{figure}[h]    \includegraphics[width=0.9\linewidth]{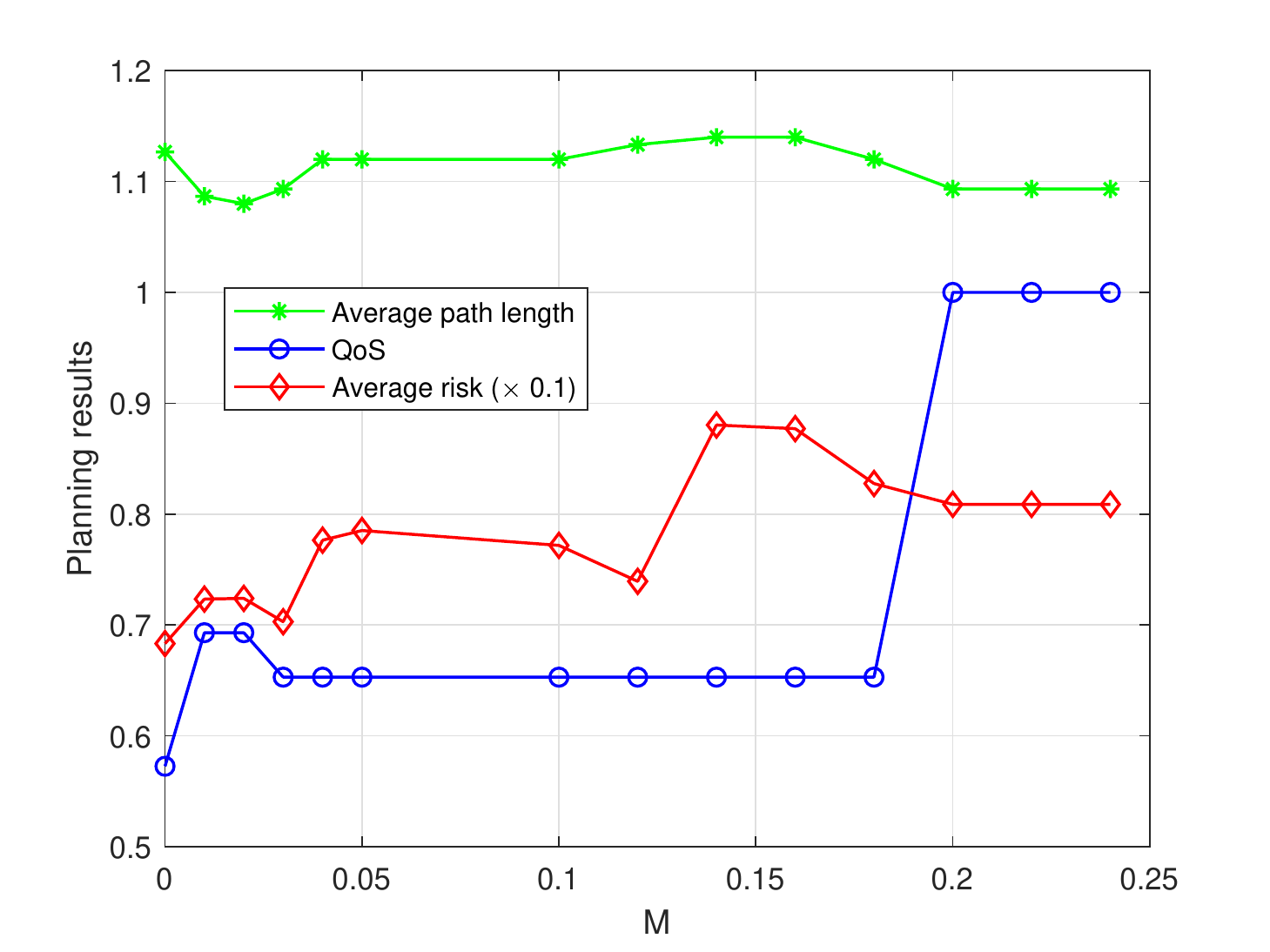}
    \caption{Planning result measurements with different $M~(K=5)$}
    \label{results_vs_M}
\end{figure}

As shown in Figure \ref{results_vs_M}, when $M$ increases, $QoS$ and average risk increase accordingly.

By changing $M$, our platform can meet the requirements of different missions with different service needs. $QoS$ and risk can be flexibly balanced.

\subsection{The Evaluation of $K$}

\begin{figure}[ht]
    \centering
    \subfigure[$K=2$]{
    \includegraphics[width=0.45\linewidth]{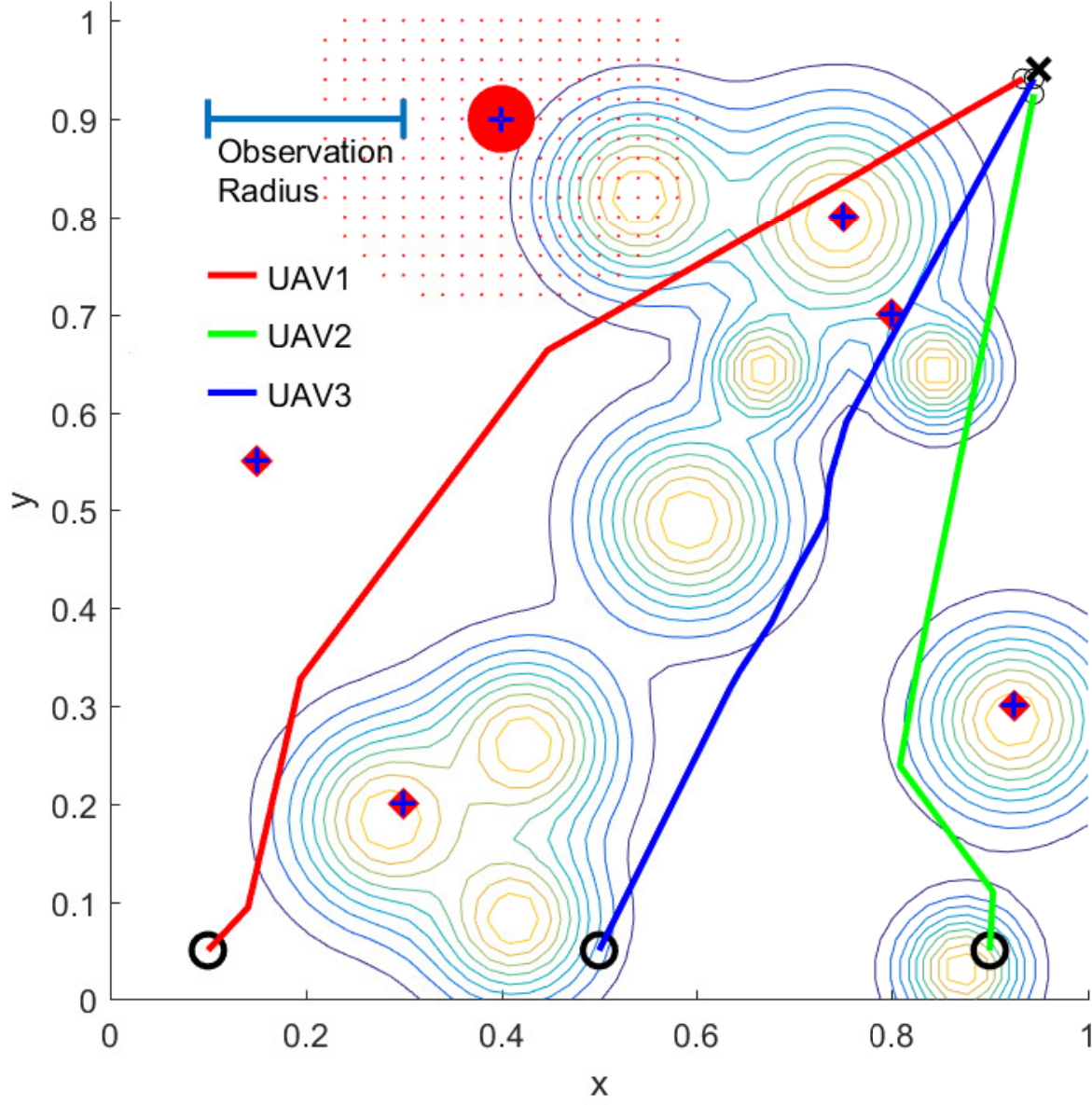}
    \label{M=0.2,K=2}}
    \subfigure[$K=5$]{
    \includegraphics[width=0.45\linewidth]{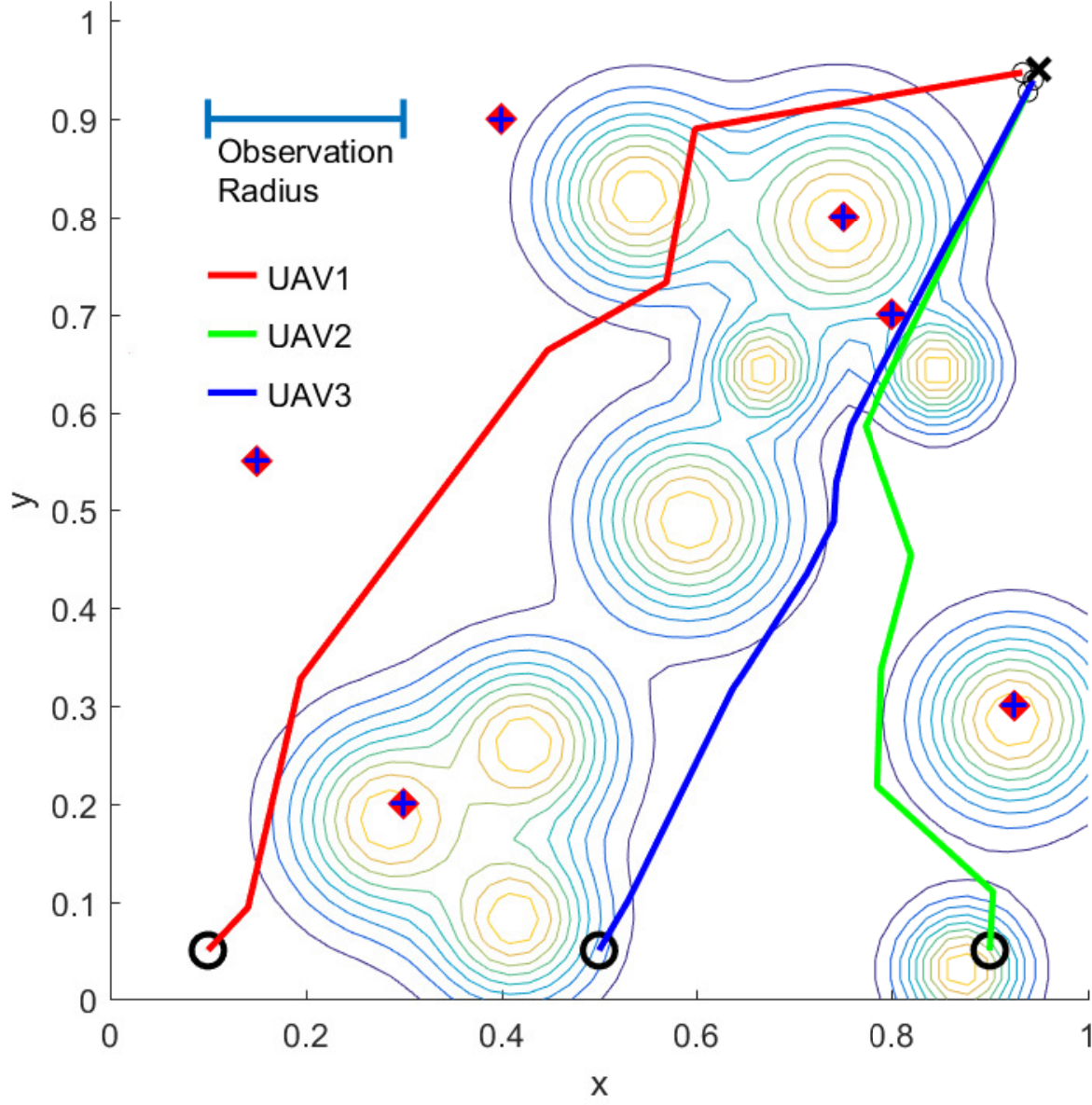}
    \label{M=0.2,K=5}}
    \subfigure[$K=10$]{
    \includegraphics[width=0.46\linewidth]{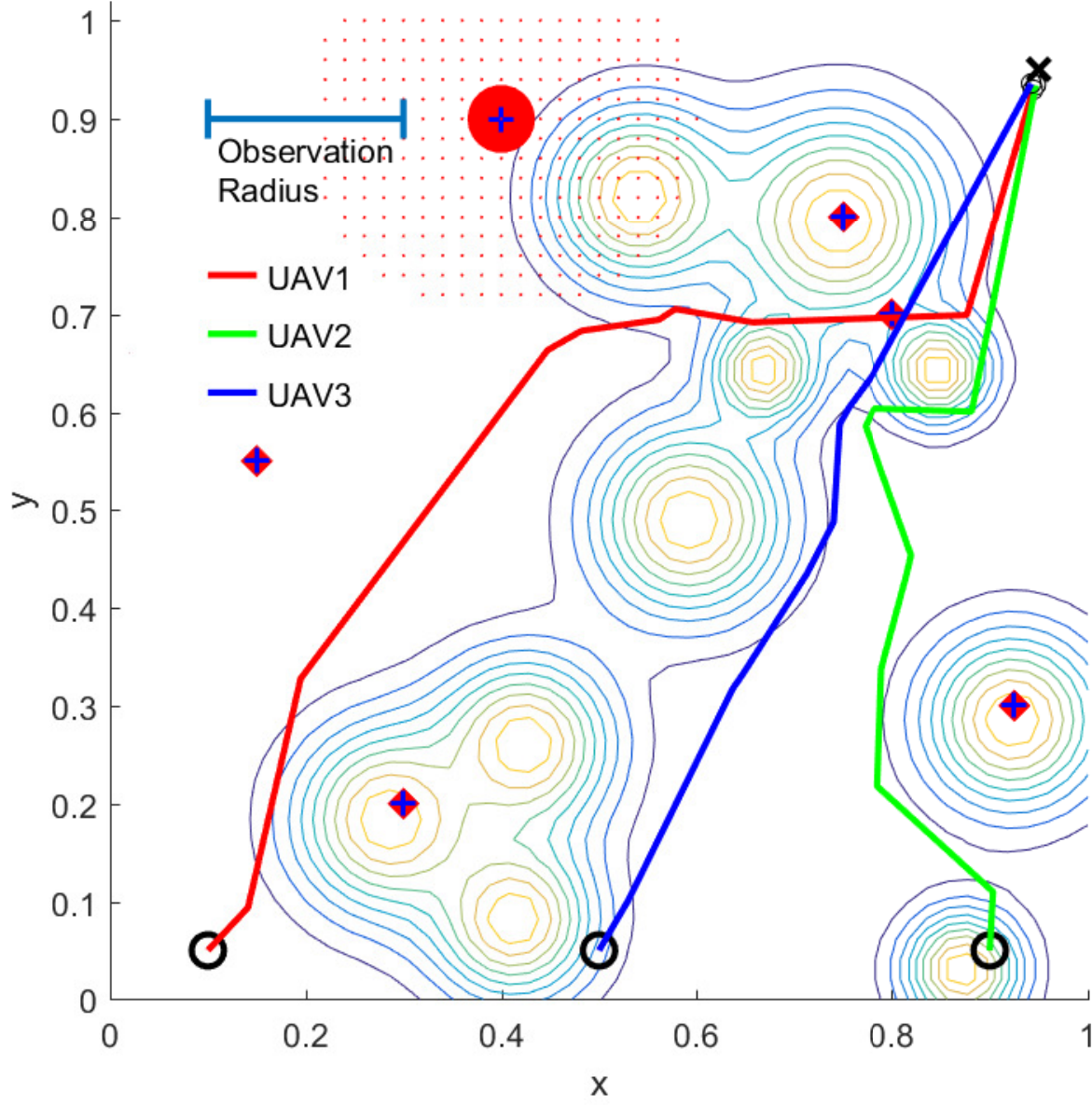}
    \label{M=0.2,K=10}}
    \subfigure[$K=50$]{
    \includegraphics[width=0.46\linewidth]{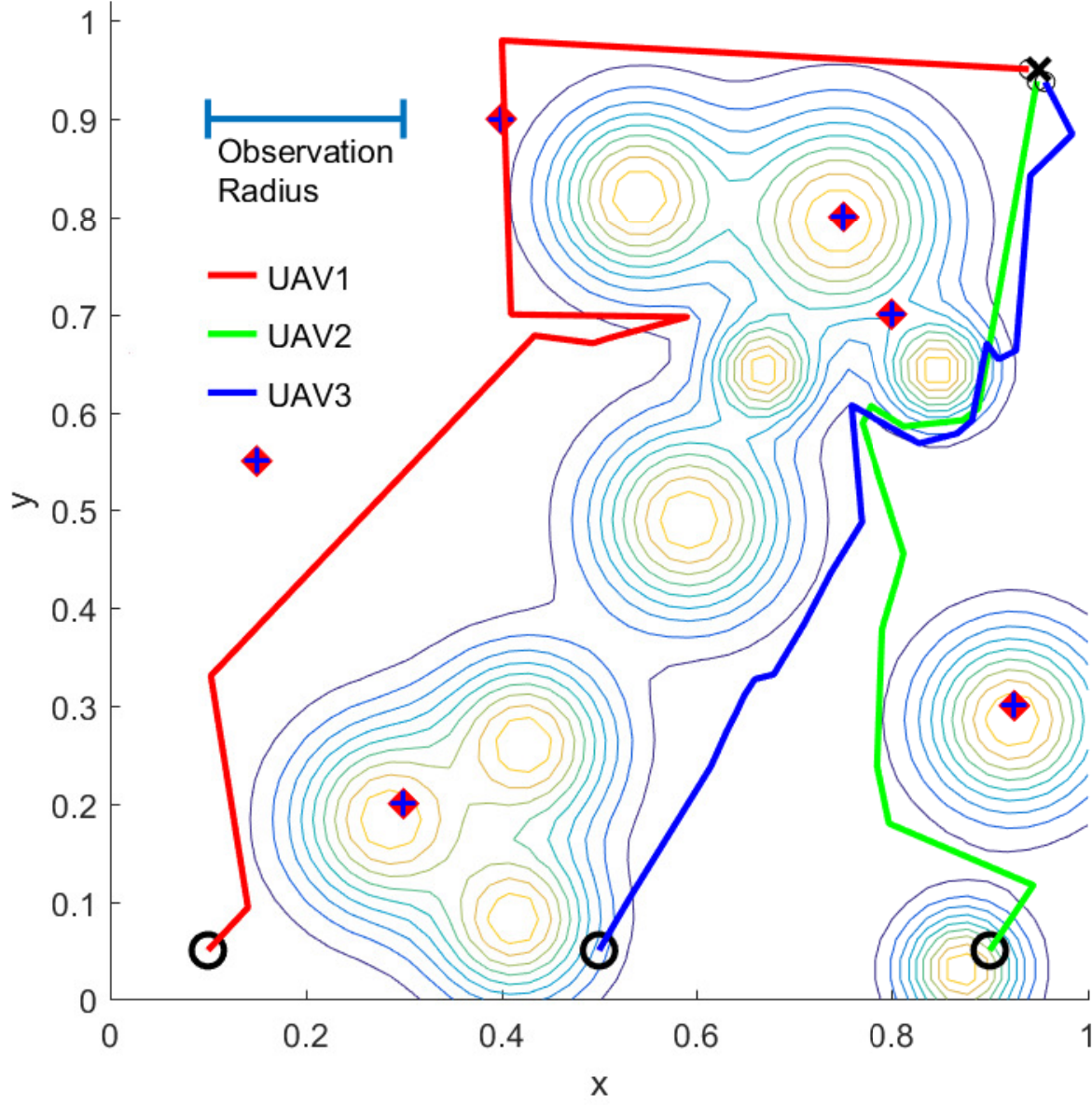}
    \label{M=0.2,K=50}}

    \caption{Planning results with different $K~(M=0.2)$}
    \label{4K}
\end{figure}

Similar to parameter $M$, altering the parameter $K$ can make the algorithm more flexible in different environments.
$K$ controls the tolerance of risk. UAVs set with higher $K$ in the result tend to sacrifice energy cost to avoid risk and thereby influence $QoS$. 
Figure \ref{4K} compares the path planning results with different $K$. When $K$ increases, a higher risk avoidance strategy is adopted, and the UAV follows a path to avoid obstacles at all costs instead of flying through narrow tunnels to serve the terminal users.

\begin{figure}[h]
    \includegraphics[width=0.9\linewidth]{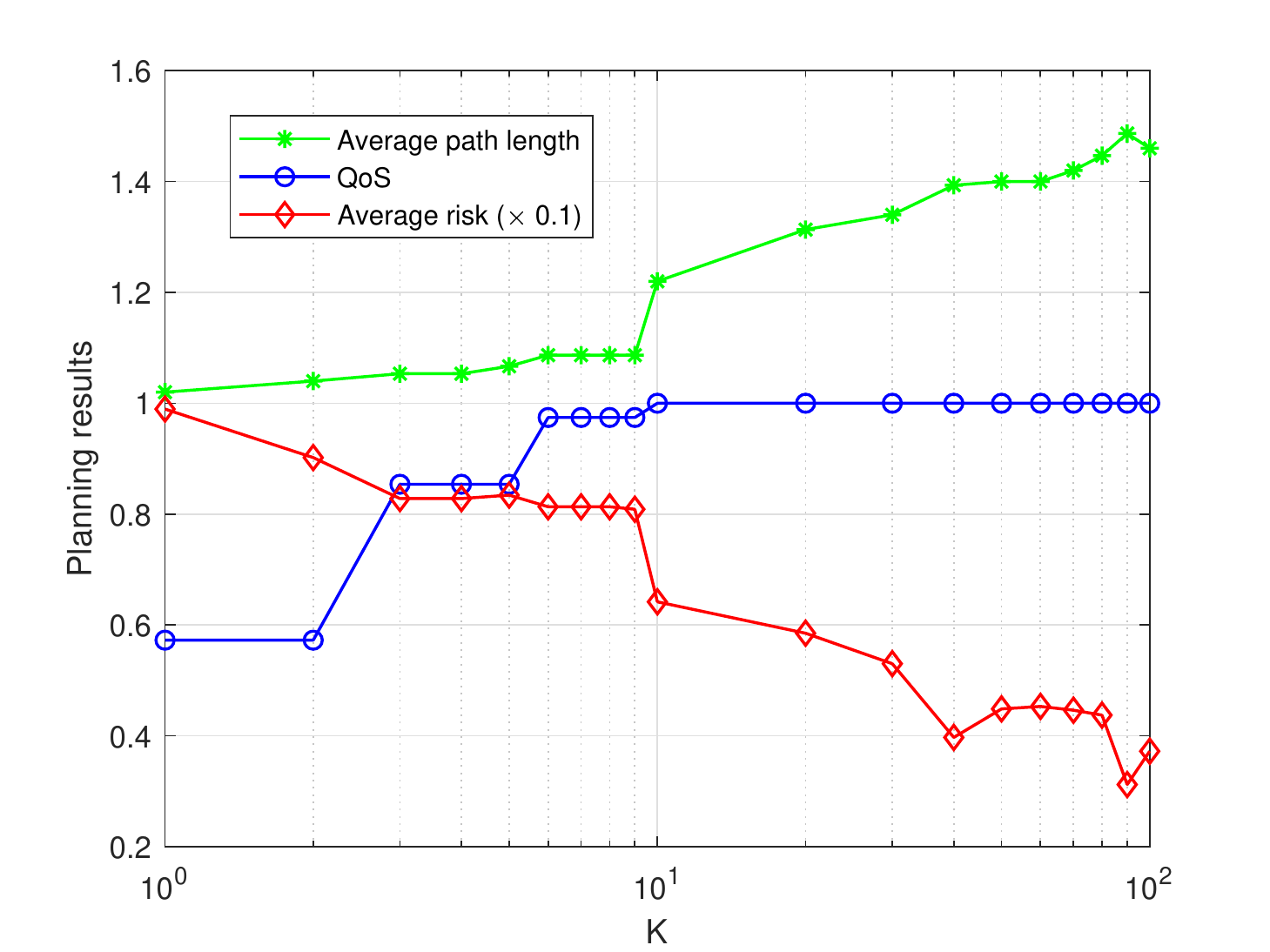}
    \caption{Planning results measurements with changing $K~(M=0.5)$}
    \label{results_vs_K}
\end{figure}

Numerical results are shown in Figure \ref{results_vs_K} to better illustrate the algorithm. When $K$ increases, path length and path risk move in opposite directions.

\subsection{Comparison of the sigmoid demand function and linear demand function}

\begin{figure}[h]
    \centering
    \includegraphics[width=0.9\linewidth]{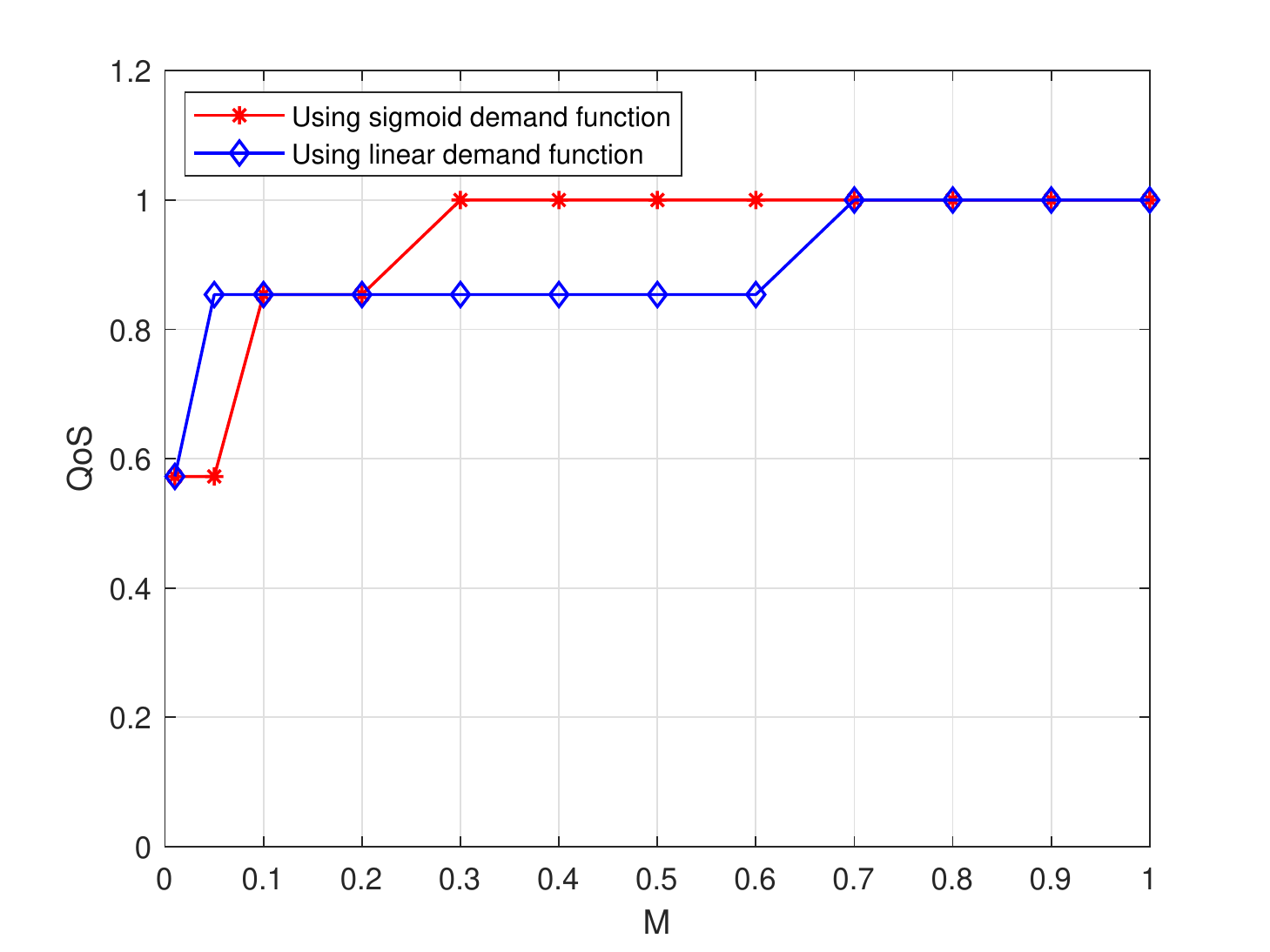}
    \caption{$QoS$ comparison using sigmoid or linear demand function ($K=10$)}
    \label{liner_sigmoid}
\end{figure}

According to \cite{liu2020path}, a sigmoid processed demand can improve system performance. To verify the effectiveness of \eqref{eq5}, we performed experiments separately with the sigmoid demand function $\sum_{j\in s(p,\epsilon)}U(d_j)$ and the linear demand function $\sum_{j\in s(p,\epsilon)}d_j$ where $d_j\in[0,1]$. Results show that the former gives higher $QoS$ with the same experimental conditions and leads to higher service speed with the same $QoS$.
Figure \ref{liner_sigmoid} shows that the sigmoid demand function assures a higher $QoS$ than the linear demand function.

Table \ref{table1} compares the service completion time of each terminal user with the two schemes. Accordingly, the sigmoid demand function leads to a higher terminal user service speed with the same $QoS$. This is because $U(d_j)>d_j$ in the early period, as shown in \eqref{eq7} and \eqref{eq8}. The UAVs are attracted to the terminal users earlier and thus finish the service more quickly.

\begin{figure}[ht]
    \centering
    \subfigure[Using linear demand function]{
    \includegraphics[width=0.45\linewidth]{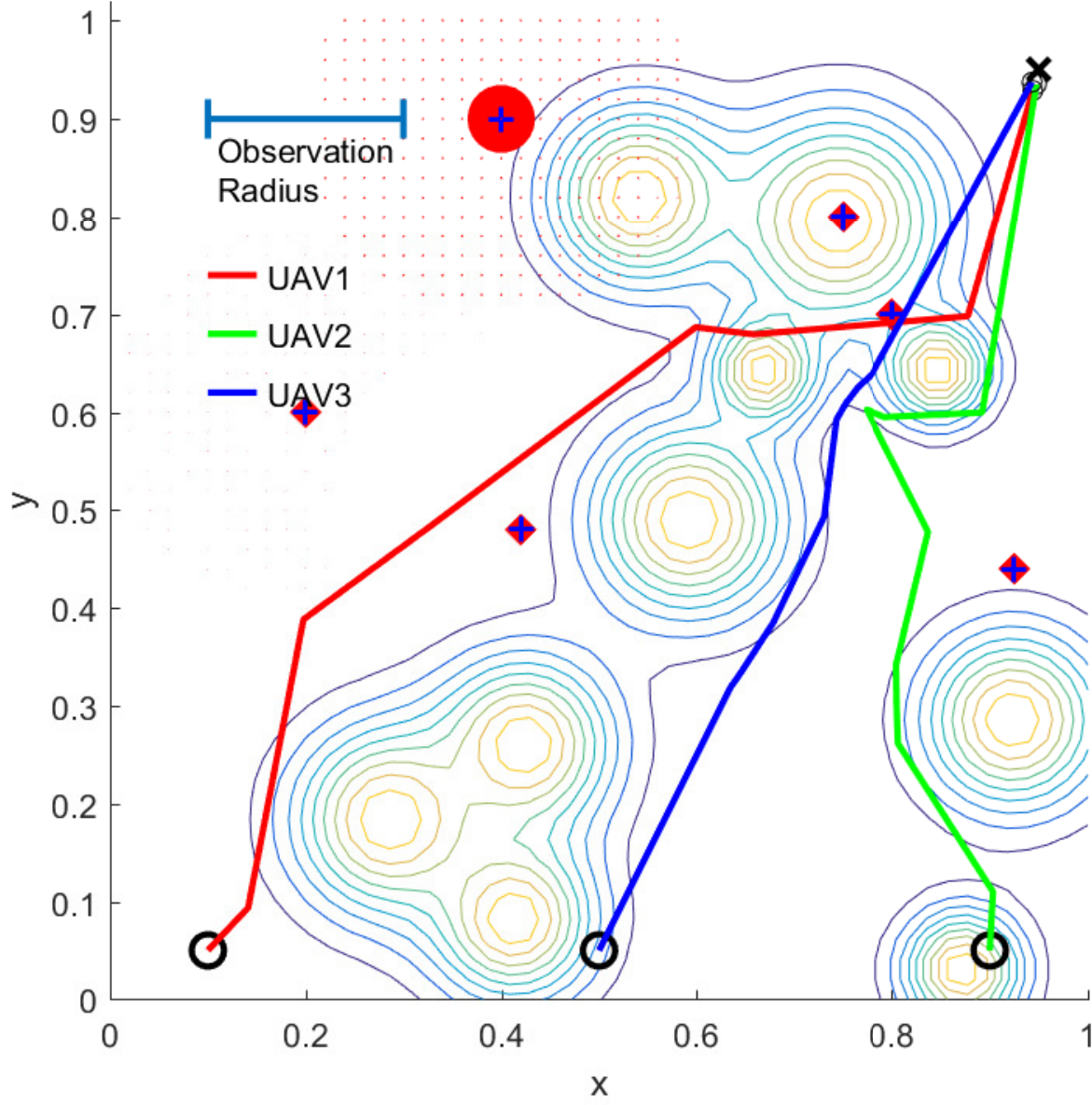}
    \label{linear_K=10_M=0.12}}
    \subfigure[Using sigmoid demand function]{
    \includegraphics[width=0.45\linewidth]{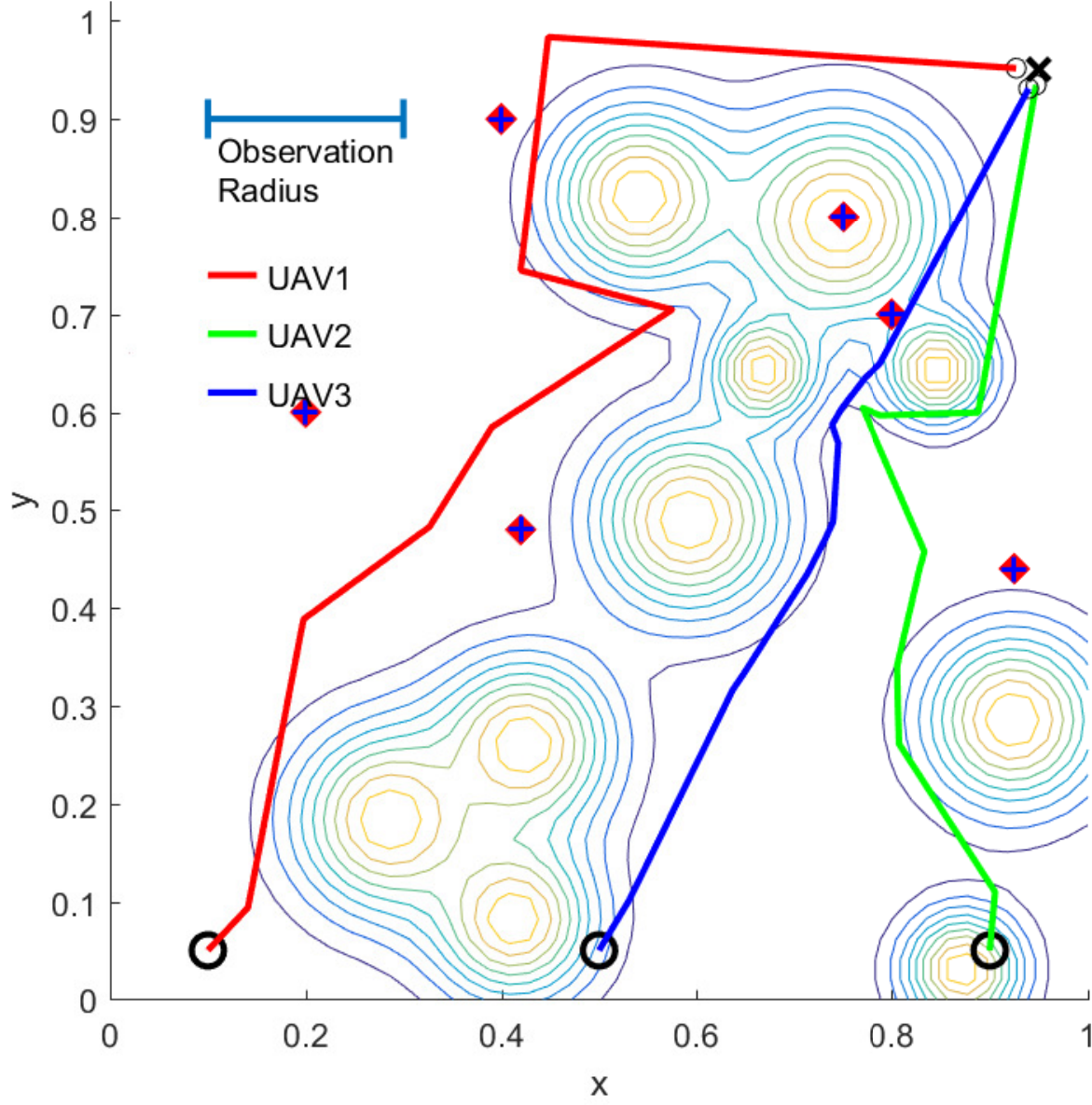}
    \label{sigmoid_K=10_M=0.12}}
    
    \caption{Results comparison using linear or sigmoid demand function ($K=10, M=0.12$)}
    \label{sigmoid_vs_linear}
\end{figure}

\begin{table*}[ht]
\centering
\caption{Service speed comparison using sigmoid or linear demand function}
\label{table1}
\begin{tabular}{|c|c|c|c|c|c|c|}
\hline
\multirow{2}{*}{$QoS$} & \multicolumn{6}{c|}{Ratio of service completion time (sigmoid vs linear)}                                 \\ \cline{2-7} 
                       & Terminal user 1 & Terminal user 2 & Terminal user 3 & Terminal user 4 & Terminal user 5 & Terminal user 6 \\ \hline
0.85 ($K=10, M=0.1$) & 1    & 1 & 1    & 1.03 & 0.82 & N    \\
0.85 ($K=10, M=0.2$) & 0.96 & 1 & 0.96 & 1    & 0.96 & N    \\
1.00 ($K=10, M=0.7$) & 1    & 1 & 1    & 1    & 0.95 & 0.96 \\ \hline
\end{tabular}
\end{table*}

\subsection{Comparison of our algorithm with the A* algorithm}

The A* algorithm is widely used as a baseline in various path planning scenarios. We compare our proposed algorithm with A* in the multiple UAV mobile edge computing environment. 

Based on the A* algorithm, at each step of path planning, a UAV can choose among one of a fixed number of equally distributed directions to move one unit step. In our experiments, we set up eight directions for UAVs and thus have eight candidate nodes $p_i$ for a UAV to choose at each step. Considering the path length and risk as a cost and the demand of the terminal users as a reward, we formulate the weight function $F_i$ at each point $p_i$ in the A* algorithm as
\begin{equation}
\begin{split}
    F_i = d_{p_i,p_t} + K R_{p_i} + \frac{M}{1+\sum_{j\in s(p_i,\epsilon)}U(d_j)},
    \\i=1,2,\cdots,7,8,
\end{split}
\label{A}
\end{equation}
where $d_{p_i,p_t}$ is the Euclidean distance between candidate point $p_i$ and the target point $p_t$.  $R_{p_i}$ and $U(d_j)$ are defined in the previous equations.

\begin{figure}[ht]
    \centering
    \subfigure[A*]{
    \includegraphics[width=0.465\linewidth]{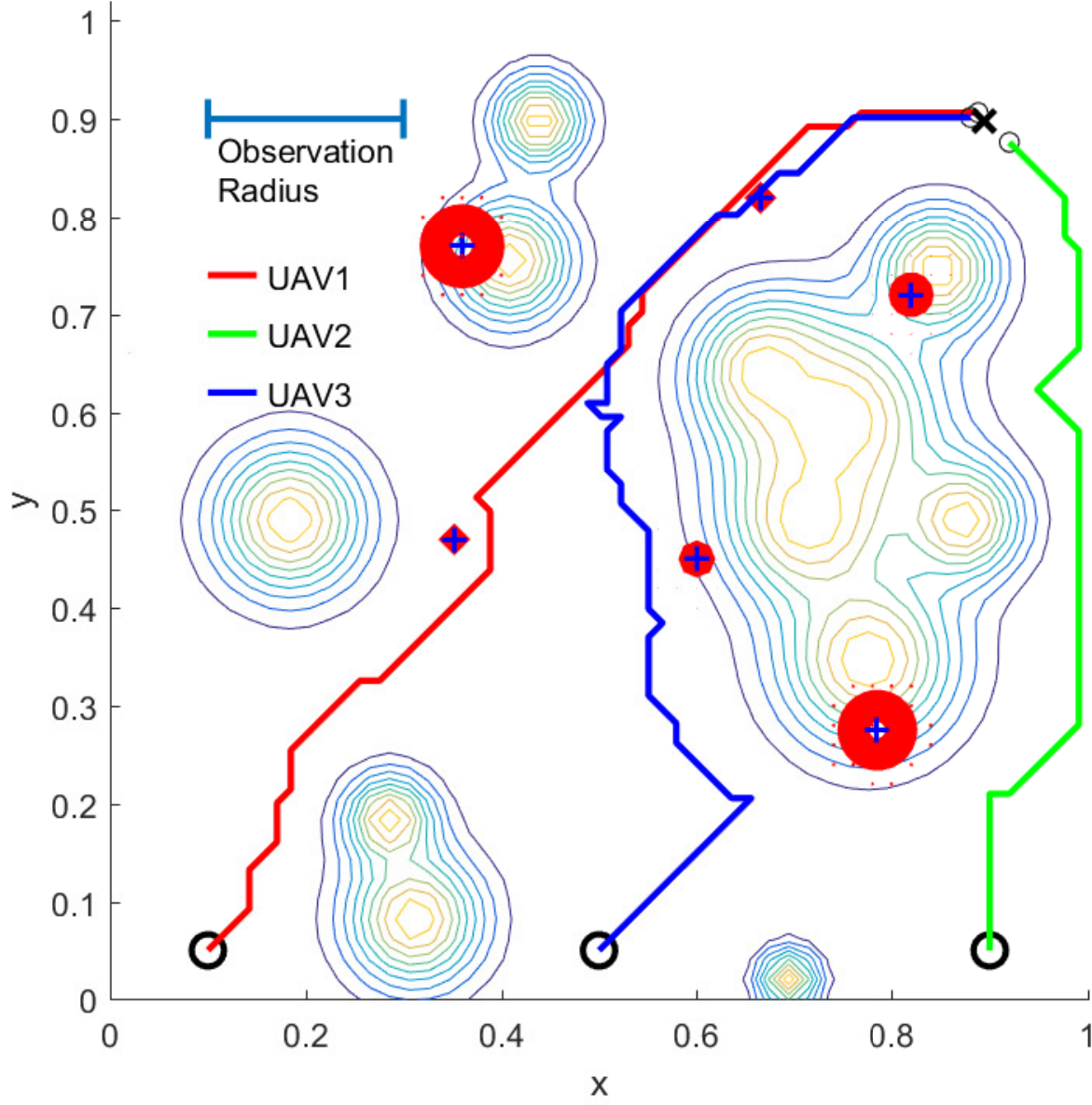}
    \label{A}}
    \subfigure[RL]{
    \includegraphics[width=0.465\linewidth]{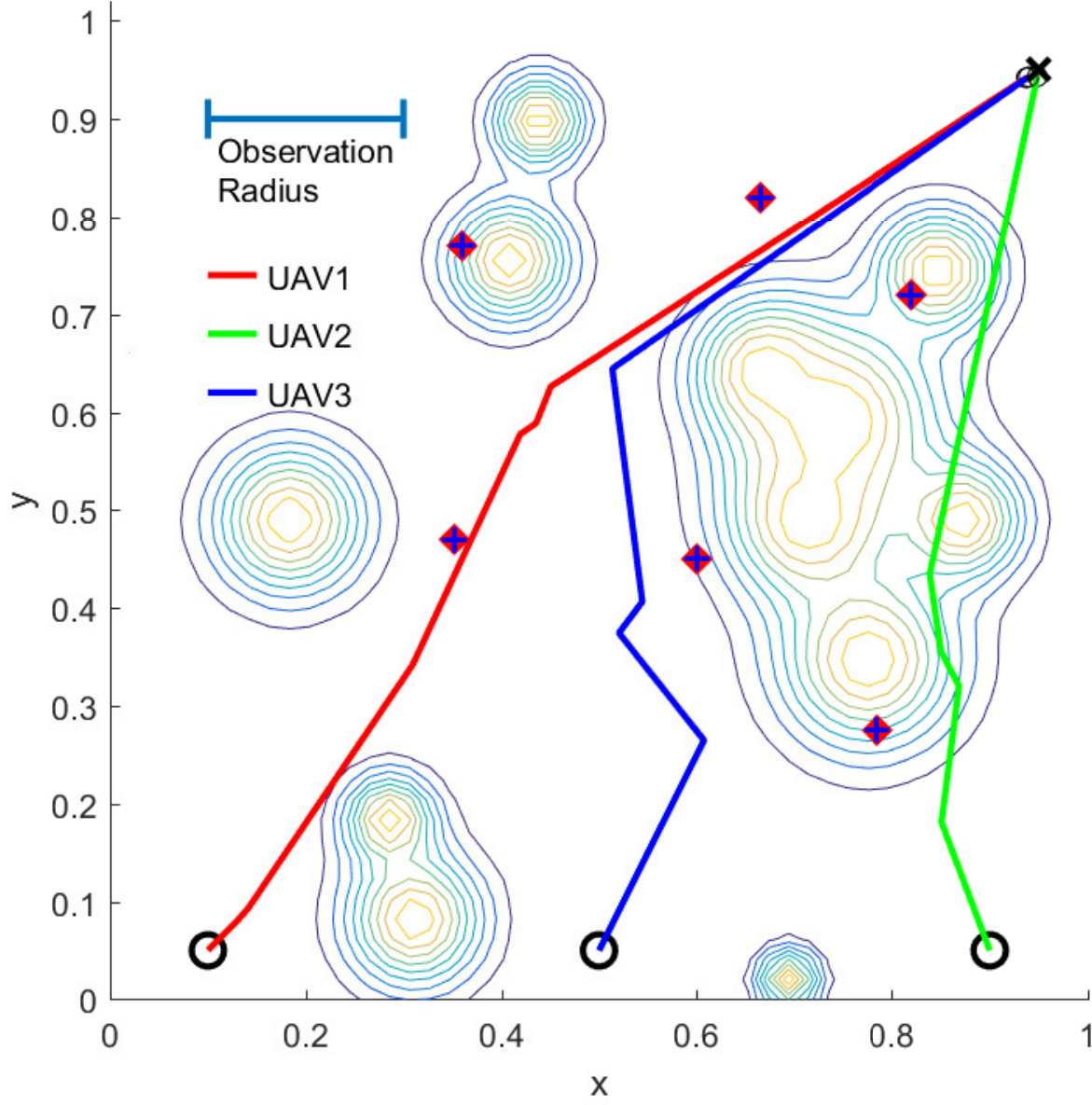}
    \label{RL}}
    \caption{Comparison of RL and A* algorithm ($K=5,M=0.5$)}
    \label{comparison with A}
\end{figure}

In general, Figure \ref{A} shows the A* algorithm fails to perform effective service when terminal users are surrounded by obstacles. However, with the proposed algorithm (Figure \ref{RL}), the UAVs manage to serve the terminal users while avoiding risk.

$QoS$ is the most important index in a mobile edge computing mission. As shown in Table \ref{table2} and Table \ref{table3} , the proposed algorithm can achieve a higher $QoS$ compared to the A* algorithm. Furthermore, our algorithm can react flexibly to the change of $K$ and $M$, thus adjust priorities to path length and average risk while assuring $QoS$. For example, with the proposed algorithm ($K=50,M=0.5$), all three indices are better than using the A* algorithm ($K=0.5,M=0.5$).

\begin{figure}[ht]
    \centering
    \includegraphics[width=0.9\linewidth]{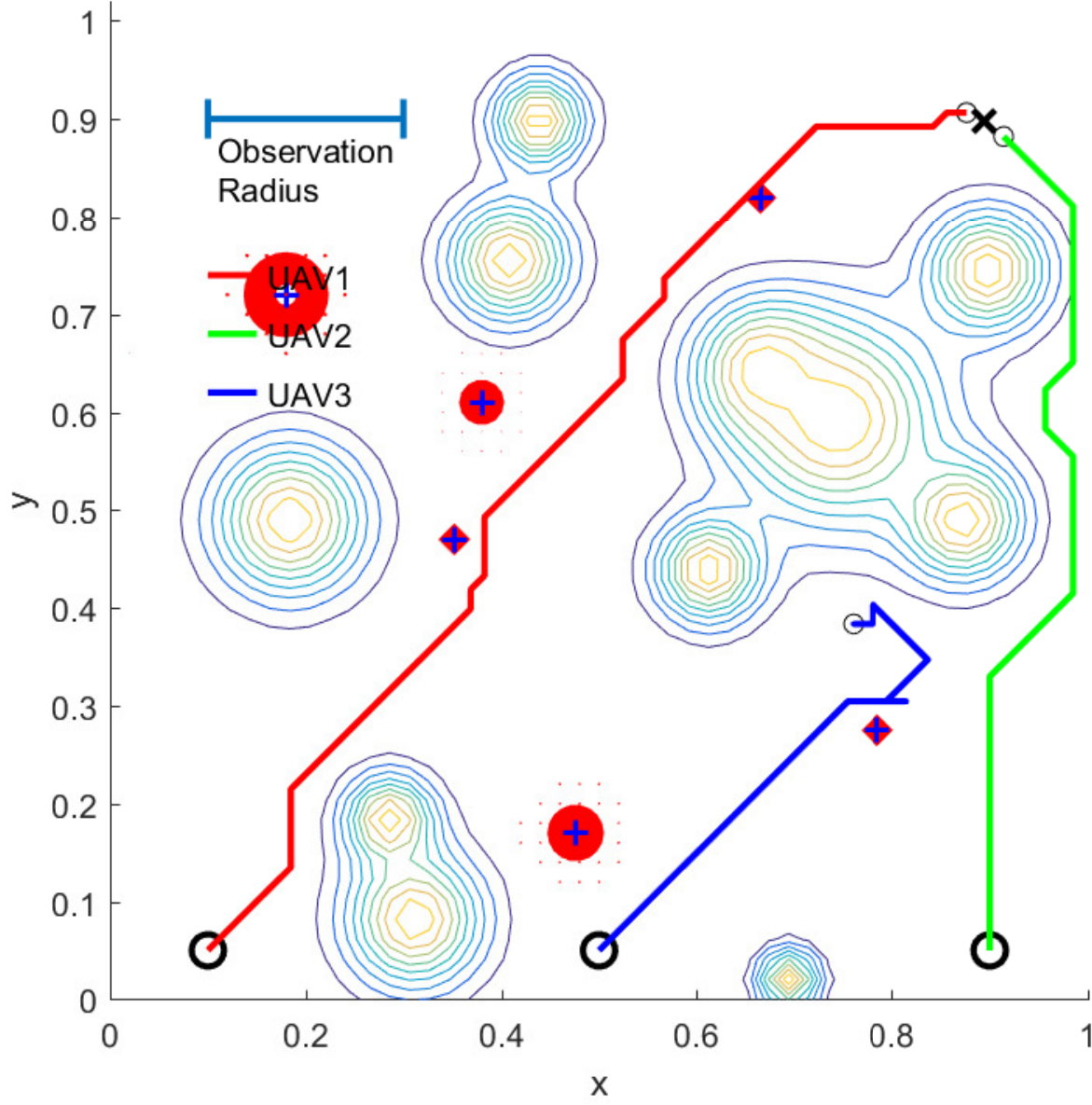}
    \caption{The deadlock of the A* algorithm ($K=1,M=0.2$)}
    \label{deadlock}
\end{figure}

A deadlock means that the mission can never come to an end because one or more UAVs are stuck in the current surrounding and loop infinitely. Our experiments show that the A* algorithm can easily fall into a deadlock with parameter $K$ above a certain level. The results are illustrated in Table \ref{table3} and Figure \ref{deadlock}.

\begin{table*}[ht]
\centering
\caption{Results comparison with different $M$ of the proposed algorithm and the A* algorithm ($K=2$)}
\label{table2}
\begin{tabular}{|c|c|c|c|c|c|c|}
\hline
\multirow{2}{*}{M} & \multicolumn{3}{c|}{Proposed algorithm} & \multicolumn{3}{c|}{A* algorithm}        \\ \cline{2-7} 
                   & $QoS$  & Average path length  & Average risk & $QoS$ & Average path length & Average risk \\ \hline
0.01 & 0.87 & 1.06 & 1.68 & 0.36 & 1.13 & 0.30 \\
0.05 & 0.85 & 1.07 & 2.00 & 0.39 & 1.13 & 0.30 \\
0.10 & 0.88 & 1.09 & 2.00 & 0.56 & 1.17 & 0.32 \\
0.15 & 0.88 & 1.09 & 2.62 & 0.56 & 1.17 & 0.32 \\
0.50 & 1.00 & 1.09 & 4.42 & 0.61 & 1.19 & 0.31 \\
1.00 & 1.00 & 1.09 & 4.33 & 0.61 & 1.19 & 0.31 \\
2.00 & 1.00 & 1.17 & 3.83 & 0.61 & 1.19 & 0.31 \\
5.00 & 1.00 & 1.17 & 4.23 & 0.61 & 1.19 & 0.31 \\
10.0 & 1.00 & 1.17 & 4.23 & 0.61 & 1.19 & 0.31 \\ \hline
\end{tabular}
\end{table*}

\begin{table*}[ht]
\centering
\caption{Results comparison with different $K$ of the proposed algorithm and A* algorithm ($M=0.5$)}
\label{table3}
\begin{tabular}{|c|c|c|c|c|c|c|}
\hline
\multirow{2}{*}{K} & \multicolumn{3}{c|}{Proposed algorithm} & \multicolumn{3}{c|}{A* algorithm}              \\ \cline{2-7} 
                   & $QoS$ & Average path length & Average risk & $QoS$   & Average path length  & Average risk  \\ \hline
0.50 & 1.00 & 1.07 & 6.98 & 0.59  & 1.13  & 0.61  \\
1.00 & 1.00 & 1.09 & 4.77 & 0.59  & 1.17  & 0.45  \\
2.00 & 1.00 & 1.09 & 4.42 & 0.61  & 1.19  & 0.31  \\
3.00 & 1.00 & 1.09 & 4.42 & 0.61  & 1.17  & 0.28  \\
5.00 & 1.00 & 1.10 & 4.54 & 0.59  & 1.19  & 0.21  \\ \cline{5-7} 
10.0               & 1.00  & 1.12                & 1.41         & \multicolumn{3}{c|}{\multirow{4}{*}{Deadlock}} \\
20.0 & 1.00 & 1.12 & 1.16 & \multicolumn{3}{c|}{} \\
50.0 & 1.00 & 1.13 & 0.42 & \multicolumn{3}{c|}{} \\
100  & 1.00 & 1.15 & 0.36 & \multicolumn{3}{c|}{} \\ \hline
\end{tabular}
\end{table*}

Based on the above comparisons, our RL platform performs better not only on environment adaptation and high $QoS$ assurance but also on algorithm reliability and high mission completeness. The effectiveness and feasibility of our method are clear.

\section{Conclusion}

This paper develops the first multi-UAV mobile edge computing and path planning platform where UAVs serve terminal users as mobile network processors based on reinforcement learning.
We implemented our platform with 1) a Quality-of-Service ($QoS$) for each terminal user, 2) maximum collision avoidance with minimum risk, and 3) cooperation between UAVs.
The simulations and experiments are provided to show the efficiency and usability of the platform, which can be a useful baseline for mobile edge computing.

\appendix
    \section{Appendix}
    \begin{proof}
        According to Property \ref{co1}, for any $x>0$ in $U(x)$,
        \begin{subnumcases}{}
            \frac{dU}{dx}=\frac{x^{\eta-1}e^{-\frac{x^\eta}{x+\beta}}}{(x+\beta)^2}[(\eta-1)x+\eta\beta]>0,\\
            1-e^{-\frac{x^\eta}{x+\beta}}>0.
        \end{subnumcases}
    
        then we have
        \begin{subnumcases}{}
            (\eta-1)x+\eta\beta>0,\label{PR1-2a}\\
            \frac{x^\eta}{x+\beta}>0.
            \label{PR1-2b}
        \end{subnumcases}
        
        According to \eqref{PR1-2b}, because $x^\eta>0$, then for any $x>0, x+\beta>0$, so $\beta>0$. By \eqref{PR1-2a}, we have $\eta>\frac{x}{x+\beta}>0$, then $\eta>0$.
    
        If $\eta<=1$, then $x<=\frac{\eta\beta}{1-\eta}$, which contradicts the domain of $x$, thus we have $\eta>1$
        \label{proof1}
    \end{proof} 

    \begin{proof}
        We have
        \begin{equation}
            \frac{dU}{d\eta}=\frac{\ln{x}e^{-\frac{x^\eta}{x+\beta}}}{x+\beta}x^\eta,
        \end{equation}
    
        thus we have 
        \begin{subnumcases}{}
            \frac{dU}{d\eta}<0, &$x\in[0,1)$,\\
            \frac{dU}{d\eta}>0, &$x>1$.
        \end{subnumcases}
    
        It means as $\eta$ increases, $U(x)$ decreases when $x<1$ and increases when $x>1$, and the slope of the curve increases accordingly. When $x=1$, $\frac{dU}{d\eta}=0$ for any $\eta$, it creates an inflection point that is not affected by changing $\eta$.
    
        We also have
        \begin{equation}
            \frac{dU}{d\beta}=-\frac{x^{\eta}e^{-\frac{x^\eta}{x+\beta}}}{(x+\beta)^2}<0,
        \end{equation}
        which means when $\beta$ increases, the sigmoid demand curve drops vertically to the inflection point.
        \label{proof2}
    \end{proof}

\section*{Acknowledgments.} The work was supported in part by National Natural
Science Foundation of China under Grant 62076016. Baochang Zhang is the correspondence.

\bibliographystyle{IEEEtran}

\bibliography{References}

\begin{thebibliography}{10}
\providecommand{\url}[1]{#1}
\csname url@samestyle\endcsname
\providecommand{\newblock}{\relax}
\providecommand{\bibinfo}[2]{#2}
\providecommand{\BIBentrySTDinterwordspacing}{\spaceskip=0pt\relax}
\providecommand{\BIBentryALTinterwordstretchfactor}{4}
\providecommand{\BIBentryALTinterwordspacing}{\spaceskip=\fontdimen2\font plus
\BIBentryALTinterwordstretchfactor\fontdimen3\font minus
  \fontdimen4\font\relax}
\providecommand{\BIBforeignlanguage}[2]{{%
\expandafter\ifx\csname l@#1\endcsname\relax
\typeout{** WARNING: IEEEtran.bst: No hyphenation pattern has been}%
\typeout{** loaded for the language `#1'. Using the pattern for}%
\typeout{** the default language instead.}%
\else
\language=\csname l@#1\endcsname
\fi
#2}}
\providecommand{\BIBdecl}{\relax}
\BIBdecl

\bibitem{abbas2017mobile}
N.~Abbas, Y.~Zhang, A.~Taherkordi, and T.~Skeie, ``Mobile edge computing: A
  survey,'' \emph{IEEE Internet of Things Journal}, vol.~5, no.~1, pp.
  450--465, 2017.

\bibitem{mao2017survey}
Y.~Mao, C.~You, J.~Zhang, K.~Huang, and K.~B. Letaief, ``A survey on mobile
  edge computing: The communication perspective,'' \emph{IEEE Communications
  Surveys \& Tutorials}, vol.~19, no.~4, pp. 2322--2358, 2017.

\bibitem{mach2017mobile}
P.~Mach and Z.~Becvar, ``Mobile edge computing: A survey on architecture and
  computation offloading,'' \emph{IEEE Communications Surveys \& Tutorials},
  vol.~19, no.~3, pp. 1628--1656, 2017.

\bibitem{du2019joint}
Y.~Du, K.~Yang, K.~Wang, G.~Zhang, Y.~Zhao, and D.~Chen, ``Joint resources and
  workflow scheduling in uav-enabled wirelessly-powered mec for iot systems,''
  \emph{IEEE Transactions on Vehicular Technology}, vol.~68, no.~10, pp.
  10\,187--10\,200, 2019.

\bibitem{zeng2016wireless}
Y.~Zeng, R.~Zhang, and T.~J. Lim, ``Wireless communications with unmanned
  aerial vehicles: Opportunities and challenges,'' \emph{IEEE Communications
  Magazine}, vol.~54, no.~5, pp. 36--42, 2016.

\bibitem{gupta2015survey}
L.~Gupta, R.~Jain, and G.~Vaszkun, ``Survey of important issues in uav
  communication networks,'' \emph{IEEE Communications Surveys \& Tutorials},
  vol.~18, no.~2, pp. 1123--1152, 2015.

\bibitem{yang2020multi}
L.~Yang, H.~Yao, J.~Wang, C.~Jiang, A.~Benslimane, and Y.~Liu, ``Multi-uav
  enabled load-balance mobile edge computing for iot networks,'' \emph{IEEE
  Internet of Things Journal}, 2020.

\bibitem{luo2020optimization}
Y.~Luo, W.~Ding, B.~Zhang, W.~Huang, and C.~Liu, ``Optimization of bits
  allocation and path planning with trajectory constraint in uav-enabled mobile
  edge computing system,'' \emph{Chinese Journal of Aeronautics}, 2020.

\bibitem{liu2019trajectory}
X.~Liu, Y.~Liu, Y.~Chen, and L.~Hanzo, ``Trajectory design and power control
  for multi-uav assisted wireless networks: A machine learning approach,''
  \emph{IEEE Transactions on Vehicular Technology}, vol.~68, no.~8, pp.
  7957--7969, 2019.

\bibitem{liu2020path}
Q.~Liu, L.~Shi, L.~Sun, J.~Li, M.~Ding, and F.~Shu, ``Path planning for
  uav-mounted mobile edge computing with deep reinforcement learning,''
  \emph{IEEE Transactions on Vehicular Technology}, vol.~69, no.~5, pp.
  5723--5728, 2020.

\bibitem{faraci2019reinforcement}
G.~Faraci, C.~Grasso, and G.~Schembra, ``Reinforcement-learning for management
  of a 5g network slice extension with uavs,'' in \emph{IEEE INFOCOM 2019-IEEE
  Conference on Computer Communications Workshops (INFOCOM WKSHPS)}.\hskip 1em
  plus 0.5em minus 0.4em\relax IEEE, 2019, pp. 732--737.

\bibitem{huang2019deep}
L.~Huang, X.~Feng, C.~Zhang, L.~Qian, and Y.~Wu, ``Deep reinforcement
  learning-based joint task offloading and bandwidth allocation for multi-user
  mobile edge computing,'' \emph{Digital Communications and Networks}, vol.~5,
  no.~1, pp. 10--17, 2019.

\bibitem{wang2019multi}
Q.~Wang, W.~Zhang, Y.~Liu, and Y.~Liu, ``Multi-uav dynamic wireless networking
  with deep reinforcement learning,'' \emph{IEEE Communications Letters},
  vol.~23, no.~12, pp. 2243--2246, 2019.

\bibitem{beck2014mobile}
M.~T. Beck, M.~Werner, S.~Feld, and S.~Schimper, ``Mobile edge computing: A
  taxonomy,'' in \emph{Proc. of the Sixth International Conference on Advances
  in Future Internet}.\hskip 1em plus 0.5em minus 0.4em\relax Citeseer, 2014,
  pp. 48--55.

\bibitem{kim2014coordinated}
S.~Kim, H.~Oh, J.~Suk, and A.~Tsourdos, ``Coordinated trajectory planning for
  efficient communication relay using multiple uavs,'' \emph{Control
  Engineering Practice}, vol.~29, pp. 42--49, 2014.

\bibitem{zhang2014cooperative}
B.~Zhang, W.~Liu, Z.~Mao, J.~Liu, and L.~Shen, ``Cooperative and geometric
  learning algorithm (cgla) for path planning of uavs with limited
  information,'' \emph{Automatica}, vol.~50, no.~3, pp. 809--820, 2014.

\bibitem{jeong2017mobile}
S.~Jeong, O.~Simeone, and J.~Kang, ``Mobile edge computing via a uav-mounted
  cloudlet: Optimization of bit allocation and path planning,'' \emph{IEEE
  Transactions on Vehicular Technology}, vol.~67, no.~3, pp. 2049--2063, 2017.

\bibitem{cao2018mobile}
X.~Cao, J.~Xu, and R.~Zhang, ``Mobile edge computing for cellular-connected
  uav: Computation offloading and trajectory optimization,'' in \emph{2018 IEEE
  19th International Workshop on Signal Processing Advances in Wireless
  Communications (SPAWC)}.\hskip 1em plus 0.5em minus 0.4em\relax IEEE, 2018,
  pp. 1--5.

\bibitem{cheng2019space}
N.~Cheng, F.~Lyu, W.~Quan, C.~Zhou, H.~He, W.~Shi, and X.~Shen,
  ``Space/aerial-assisted computing offloading for iot applications: A
  learning-based approach,'' \emph{IEEE Journal on Selected Areas in
  Communications}, vol.~37, no.~5, pp. 1117--1129, 2019.

\bibitem{yin2019intelligent}
S.~Yin, S.~Zhao, Y.~Zhao, and F.~R. Yu, ``Intelligent trajectory design in
  uav-aided communications with reinforcement learning,'' \emph{IEEE
  Transactions on Vehicular Technology}, vol.~68, no.~8, pp. 8227--8231, 2019.

\bibitem{lee2005non}
J.-W. Lee, R.~R. Mazumdar, and N.~B. Shroff, ``Non-convex optimization and rate
  control for multi-class services in the internet,'' \emph{IEEE/ACM
  transactions on networking}, vol.~13, no.~4, pp. 827--840, 2005.

\end{thebibliography}

\begin{IEEEbiography}[{\includegraphics[width=1in,height=1.25in,clip,keepaspectratio]{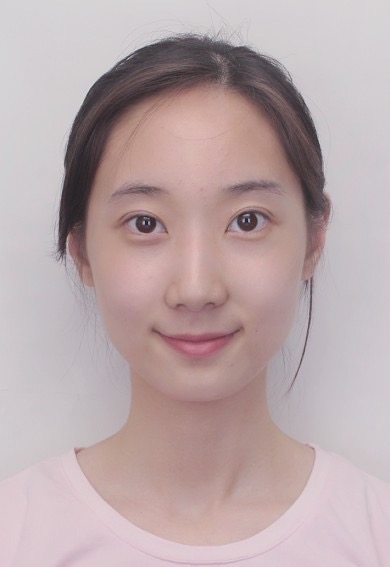}}]{Huan Chang} received her bachelor's degree in Management Science and Engineering from Beihang University, China, in 2020. She is a Master's student in the civil engineering department of Empire College London. Her study interests include RL-based path planning and optimization, supply chain management, and operational research.
\end{IEEEbiography}

\begin{IEEEbiography}[{\includegraphics[width=1in,height=1.25in,clip,keepaspectratio]{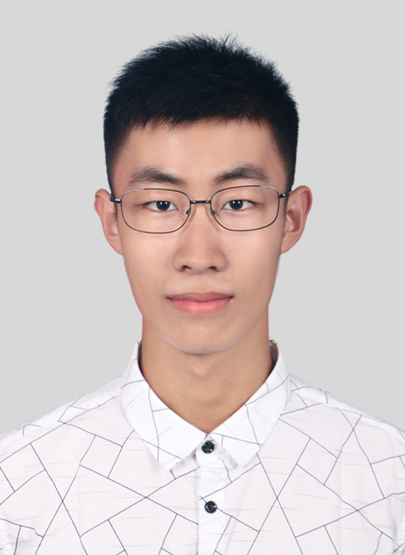}}]{Yicheng Chen} is currently working toward a B.E. degree in Automation with the School of Automation Science and Electrical Engineering, Beihang University, Beijing, China. His research interests include motion planning, pattern recognition, and reinforcement learning.
\end{IEEEbiography}

\begin{IEEEbiography}
[{\includegraphics[width=1in,height=1.25in,clip,keepaspectratio]{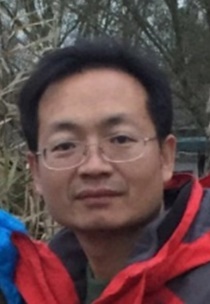}}]{Baochang Zhang} received the B.S., M.S., and Ph.D. degrees in Computer Science from Harbin Institute of the Technology, Harbin, China, in 1999, 2001, and 2006, respectively. From 2006 to 2008, he was a research fellow at the Chinese University of Hong Kong, Hong Kong, and at Griffith University, Brisbane, Australia. He was also a senior PostDoc fellow at the Italy Institute of Technology from 2014 to 2015. Currently, he is a professor at Beihang University, Beijing, China. His current research interests include deep learning, pattern recognition, object recognition and tracking, and wavelets.
\end{IEEEbiography}
   
\begin{IEEEbiography}
[{\includegraphics[width=1in,height=1.25in,clip,keepaspectratio]{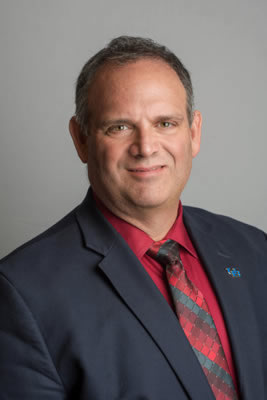}}]{David Doermann} is a Professor of Empire Innovation at the University at Buffalo (UB) and the Director of the University at Buffalo Artificial Intelligence Institute. Prior to coming to UB, he was a program manager at the Defense Advanced Research Projects Agency (DARPA), where he developed, selected, and oversaw research and transition funding in  computer vision, human language technologies, and voice analytics. From 1993 to 2018, David was a member of the research faculty at the University of Maryland, College Park. David has over 250 publications in conferences and journals, is a fellow of the IEEE and IAPR, has numerous awards, including an honorary doctorate from the University of Oulu, Finland, and is a founding Editor-in-Chief of the International Journal on Document Analysis and Recognition.
\end{IEEEbiography}
   
\end{document}